\documentclass[12pt]{article}
\usepackage{euler,epsfig,times}
\begin{document}


\newcommand{\nc}[2]{\newcommand{#1}{#2}}
\newcommand{\plaat}[3]{\raisebox{#3cm}{\epsfig{figure=#1.eps,width=#2cm}}}
\nc{\bq}{\begin{equation}}
\nc{\eq}{\end{equation}}
\nc{\bqa}{\begin{eqnarray}}
\nc{\eqa}{\end{eqnarray}}
\nc{\nl}{\nonumber \\}
\nc{\suml}{\sum\limits}
\nc{\prol}{\prod\limits}
\nc{\nsum}{\hat{\suml_{\vn}}}
\nc{\intl}{\int\limits}
\nc{\vx}{\vec{x}}
\nc{\vn}{\vec{n}}
\nc{\si}{\sigma}
\nc{\sivn}{\sigma^2_{\vn}}
\nc{\en}{e_{\vn}}
\nc{\ben}{\bar{e}_{\vn}}
\nc{\be}{\beta}
\newcommand{\fall}[2]{{#1}^{\underline{#2}}}
\nc{\der}{\partial}
\nc{\intii}{\intl_{-i\infty}^{+i\infty}}
\nc{\hz}{\hat{z}}
\nc{\mc}{Mon\-te Car\-lo}
\nc{\qmc}{Qua\-si\--Mon\-te Car\-lo}
\nc{\ovn}{\omega_{\vn}}
\newcommand{\avg}[1]{\left\langle #1\right\rangle}
\newcommand{\vari}[1]{\si\left(#1\right)^2}
\newcommand{\avgq}[1]{\left\langle #1\right\rangle_{(q)}}
\newcommand{\variq}[1]{\si\left(#1\right)^2_{(q)}}
\nc{\al}{\alpha}

\begin{center}
\begin{bf}
\begin{Large}
Error in \mc, quasi-error in \\ \qmc\
\end{Large}\\
\vspace*{\baselineskip}
Ronald Kleiss\footnote{{\tt R.Kleiss@science.ru.nl}} and
Achilleas Lazopoulos\footnote{{\tt A.Lazopoulos@science.ru.nl}}\\
\vspace*{\baselineskip}
IMAPP\\
Institute of Mathematics, Astrophysics and Particle Physics\\
Radboud University, Nijmegen, the Netherlands
\end{bf}
\end{center}

\begin{abstract}
While the \qmc\ method of numerical integration achieves smaller
integration error than standard Monte Carlo, its use in particle physics 
phenomenology has been hindered by the abscence of a 
reliable way to estimate that error. The standard \mc\ error estimator relies on the assumption that 
the points are generated independently of each other and, therefore, fails to account for the error 
improvement advertised by the \qmc\ method. We advocate the construction of an estimator of 
stochastic nature, based on the ensemble of pointsets with a particular discrepancy value. We
investigate the consequences of this choice and give some first empirical results on the 
suggested estimators. 
\end{abstract}                                          
\section{\mc\ and \qmc}
\subsection{Introduction}
In numerical integration, the main problem is not to obtain a numerical
answer\footnote{which is known to be 42, see \cite{Hitchhiker}.} 							
for the integral,
but rather, on the one hand, to ensure that the
inherent numerical error is as small as possible, and, on the other hand,
to estimate this error as precisely as possible. For integrands with well-known
smoothness properties, {\it a-priori\/} estimates of the numerical error
are possible, but for most practical applications the smoothness properties
of the integrand can only be investigated in the course of the integration
itself, that is, by repeated numerical evaluation of the integrand.

In this paper, we shall be concerned with the integration errors arising
in \mc\ and \qmc\ integration. In these methods, the
integration nodes are distributed in a (more or less) stochastic manner,
and the integration errors are therefore of an essentially probabilistic nature.
The difference between \mc\ and \qmc\ is that in the former, the integration
points are iid\footnote{iid stands for `independent, identically
distributed'.} uniform in the integration region\footnote{This ignores the
possible interpretation of stratified and importance sampling methods of
variance reduction. These can, at any rate, always be formulated in terms
of methods using iid uniform integration points.}, while in the latter
the integration points are not chosen independently, but rather with an
explicit interdependence so that their overall distribution is `smoother', in a
sense discussed below.\\

In stochastic integration methods of the \mc\ or \qmc\ types, the integration
error is itself an estimate, which contains its own error. That this is
not an academic point becomes clear when we realize that the error estimate
is routinely used to provide {\em confidence levels\/} for the integral
estimate (be it based either on Chebyshev or Central-Limit-Theorem, Gaussian
rules); and a mis-estimate of the integration error can lead to a
serious under- or overestimate of the confidence level.
As an example, suppose that the Central Limit Theorem is applicable, so that the
integration result is drawn from a Gaussian distribution centered
around the true integral value. One standard deviation, as estimated
by Monte Carlo, corresponds to a two-sided confidence level of 68\%. If the
error estimate is off by 50\% (admittedly a large value), the
actual confidence level may then be anything between 38\% and 87\%.\\

From this consideration, we are therefore
led to a hierarchy of error estimates: the {\em first-order\/} error
is that on the integral estimate, while the {\em second-order\/} error
is the error on the error estimate. This in turn has, of course, its own
{\em third-order\/} error, and so on. Higher orders than the second one,
however, appear to be too academic for practical relevance, but we should
like to argue that, in any serious integration problem, the second-order
error ought to be included. In what follows we shall discuss the first- and
second-order error estimates.\\

Due to the absence of a \qmc\ error estimator, users of \qmc\ 
have been estimating the integration error with the classical \mc\ formula, 
as if the point set was iid. This systematically overestimates the error in any case where the 
quasi point-set is of any worth. Moreover, no confidence levels can be assigned
since the classical estimator does not average to the error made 
by the quasi, non-iid point-sequence. The purpose of this paper is to investigate 
possible estimators for \qmc\ integration taking under consideration the non-iid nature
of the underlying point-set\footnote{The opposite direction - 
re-introducing randomness by reshuffling the points of the \qmc\ sequence
in a way that preserves their uniformity properties, 
thus allowing for the use of a `classical'-type estimator -  
has been studied extensively in the literature (see \cite{Owen}
and references therein). Such point-sequences behave better than \mc\
sequences and, for integrands with certain properties, as good as
\qmc\ sequences. Estimating the error, however, requires the use of
a number $r$ of different reshufflings of a point-set with $n$ points, 
thereby trading off accuracy for knowledge of the error.}.

\subsection{\mc\ estimators}
In this section we briefly review the probabilistic theory underlying
\mc\ integration. This is of course well known, but we include it here
so that the significant difference with \qmc\ can become clear.

Throughout this paper we shall consider integration problems over the
$d$-di\-men\-si\-on\-al unit hypercube $C=[0,1)^d$. The integrand is a
function $f(\vx)$, which we shall assume real and non-negative, and, of course,
integrable over $C$.
We shall define
\bq
J_m = \intl_C\;f(\vx)^m\;d^d\vx\;\;\;,\;\;\;m=1,2,3,\ldots\;\;,
\eq
so that $J_1$ is the required integral. Note that $J_m$ is not
necessarily finite for $m\ge2$.
In \mc\ we assume $N$ integration points, to be
chosen iid from the uniform probability distribution over $C$. This means that
the {\em point set\/} $X=\{\vx_1,\vx_2,\ldots,\vx_N\}$ on which the
integration is based is {\em assumed\/} to be a typical member of
an ensemble of such point sets, in such a way that the combined
probability distribution of the $N$ points over this ensemble is the uniform
iid one:
\bq
P_N(\vx_1,\vx_2,\ldots,\vx_N) = 1\;\;.\label{regulariid}
\eq
We shall take the averages over this ensemble.
Other assumptions on the underlying ensemble from which the point set $X$
is believed to be chosen are possible, leading to a different form of $P_N$.
In this, the situation is not different from that encountered in statistical
mechanics. The above assumption, however, is the one that is always made
in regular \mc\, and is justified to some extent by the fact that good-quality
(pseudo)random number generators are actually available,
allowing us to build ensembles of point sets
$X$ that indeed have the above property (\ref{regulariid}).\\

Let us assume that a point set $X$ has been generated, and the values of the
integrand $f(\vx)$ at all these points have been computed. These we shall
denote by $f_j\equiv f(\vx_j)$, $j=1,2,\ldots,N$. From these we can compute
the discrete analogues of the integrals $J_m$, which are computable in
linear time (that is, time proportional to $N$):
\bq
S_m = \suml_{j=1}^N\;(f_j)^m\;\;.
\eq
The \mc\ estimate of the integral is then
\bq
E_1 = {1\over N}S_1\;\;.
\eq
The expected value of $E_1$ over the above ensemble of point sets is
given by
\bq
\avg{E_1} = {1\over N}\suml_i\avg{f_i} = \intl_C f(\vx)\;d^d\vx = J_1\;\;,
\eq
which is indeed the required integral: this is the basis for the \mc\ method.
Its usefulness appears if we compute the variance of $E_1$:
\bq
\vari{E_1} = \avg{E_1^2} - \avg{E_1}^2 =
{1\over N}\left(J_2-J_1^2\right)\;\;.
\eq
Since this decreases as $N^{-1}$, the \mc\ method actually converges for
large $N$. Note that the leading, ${\cal O}(N^0)$, terms of
$\langle E_1^2\rangle$
and $\langle E_1\rangle^2$ cancel against each other: this is a regular
phenomenon in variance estimates of this kind\footnote{It should be
pointed out that what we estimate is the average of the squared error,
rather than the error itself, and squaring and averaging do {\em not\/}
commute. In fact, this is another reason why the second-order estimate is
relevant.}. The variance $\vari{E_1}$ is estimated by the first-order
error estimator (also called `classical' or `pseudo' estimator in what follows)
\bq
E_2 = {1\over N^2}S_2 - {1\over N^3}S_1^2\;\;,
\label{classicalestimator}\eq
for which we have
\bq
\avg{E_2} = \vari{E_1} + {\cal O}(N^{-2})\;\;.
\eq
Since $N$ is usually quite large, at least 10,000 or so, we feel justified
in working only to leading order in $N$. The squared error of $E_2$
is computed to be, to leading order in $N$,
\bq
\vari{E_2} = {1\over N^3}\left(
J_4 - 4J_3J_1 - J_2^2 + 8J_2J_1^2 - 4J_1^4\right)\;\;,
\eq
for which the estimator is
\bqa
E_4 &=&
{1\over N^7}\left(
N^3S_4 - 4N^2S_3S_1 - N^2S_2^2 + 8NS_2S_1^2 - 4S_1^4\right)\;\;.
\eqa
which can also be computed in linear time; we have
\bq
\avg{E_4} = \vari{E_2} + {\cal O}(N^{-4})\;\;.
\eq
Some details on the computation of leading-order expectation values of this
type, as well as (for purposes of illustration)
the form of the third- and fourth-order error estimators, $E_8$ and $E_{16}$,
respectively, are given in the Appendix.\\

A final remark is in order here. The Central Limit theorem, which ensures
that the error estimate can be used to derive {\em Gaussian\/} confidence
levels, can also be inferred from the computation of the higher cumulants
of the error distribution: we find for the skewness
\bq
\avg{\left(E_1-\avg{E_1}\right)^3} =
{1\over N^2}\left(J_3-3J_2J_1+2J_1^3\right)\;\;,
\eq
and the unnormalized kurtosis:
\bq
\avg{\left(E_1-\avg{E_1}\right)^4} - 3\vari{E_1} =
{1\over N^3}\left(J_4-4J_3J_1-3J_2^2+12J_2J_1^2-6J_1^4\right)\;\;,
\eq
which indicate that the higher cumulants decrease faster than the
variance with increasing $N$; we shall examine this later on for the case of
\qmc.
\subsection{\qmc\ estimators}   
\subsubsection{Multi-point distribution and correlation functions}
In contrast to the case of regular \mc, the technique of \qmc\ relies on
point sets in which the points are {\em not\/} chosen iid from the uniform
distribution, but rather interdependently. To make this more specific,
let us consider a point set $X$ of $N$ points. For each such a point set,
we may define a {\em measure of non-uniformity\/}, called a {\em discrepancy\/}
or, as in this paper, a {\em diaphony}. Its precise definition is presented
below: for now, suffice it to demand that there exist a function $D(X)$ of
the point set, which increases with its non-uniformity: $D(X)=0$ if the
point set is perfectly uniform in all possible respects, an ideal situation
that can never be obtained for any finite point set. The \qmc\ method
consists of using point sets $X$ for which $D(X)$
has some value $s$ which is (very much) smaller
than $\avg{s}$, the value that may be expected for truly iid uniform ones.

Given that such `quasi-random' point sets can be obtained, how does one use them
in numerical integration? The obvious issue here is to determine of what
ensemble the quasi-random point set $X$ can be considered to be a `typical'
member. In this paper, we should like to advocate the viewpoint that,
since the main additional property of the quasi-random point set that
distinguishes it from truly random point sets is its `anomalously small'
discrepancy $D$, the ensemble ought to consist of those point sets
that are iid uniformly, with the additional constraint that the discrepancy
$D$ has the particular value $D(X)=s$ for the actually used point
set\footnote{We do not examine the possible
alternative that the point sets in the ensemble must have discrepancy
{\em in the neighborhood} of the observed value $s$; this amounts to
the distinction between the micro-canonical and  the canonical
ensemble in statistical mechanics.}. On this premise, the \qmc\ analogue
of Eq.(\ref{regulariid}) would then be the assumption
\bq
P_N(s;\vx_1,\vx_2,\ldots,\vx_N) =
{1\over H(s)}\delta(D(X)-s)\;\;,
\eq
where $s$ is, again, the observed value of the discrepancy of $X$,
on which $P_N$ must now of course depend; and
$H(s)$ is the probability density to happen upon a point sets $X$ with this
discrepancy in the regular-\mc\ ensemble:
\bq
H(s) = \intl_C\;\delta(D(X)-s)\;d^d\vx_1\;d^d\vx_2\;d^d\vx_N\nl
\eq
The actual computation of $H(s)$ for given definition of the discrepancy is
referred to the next section. What interests us here is the fact that
$P_N$ is now no longer simply unity, since that would imply independence
of the points in the point set. Let us therefore write the
{\em multi-pont distribution\/} as
\bq
P_N(s;\vx_1,\vx_2,\ldots,\vx_N) = 1 - {1\over N}
F_2(s;\vx_1,\vx_2,\ldots,\vx_N)\;\;,
\label{multipointdistribution}\eq
where we have anticipated a factor $1/N$
in the {\em multi-point correlation\/} $F$.

\subsubsection{Properties of the correlation function}

Since the value of the
discrepancy of a given point-set $X$, should be independent of the order in which the points are generated,
$F_k(s;\vx_1\ldots \vx_k)$ must be totally symmetric; moreover, we must have
\bq
F_k(s;\vx_1,\vx_2,\ldots,\vx_K) = \intl_C\;
F_{k+1}(s;\vx_1,\vx_2,\ldots,\vx_k,\vx_{k+1})\;d^d\vx_{k+1}\;\;,
\eq
which is not as trivial as it might seem since the value of the discrepancy,
$s$, is based on the full $N$ points and not on the smaller set of $k$ or $k+1$
points. Finally, for the \qmc\ integral to
be unbiased, we must have
\bq
P_1(s;\vx_1) = 1\;\;,
\eq
so that
\bq
\intl_C\;F_2(s;\vx_1,\vx_2)\;d^d\vx_2 = 0\;\;.
\label{vanishing_property}
\eq
These remain, of course, to be proven and we shall do so in the next
section, for a particular choice of discrepancy. Moreover, we shall show
there that the multi-point correlation $F_N$ is, to leading order in $1/N$,
made up from two-point correlations $F_2$:
\bq
F_k(s;\vx_1,\vx_2,\ldots,\vx_k) = \suml_{1\le m < n\le k}
F_2(s;\vx_m,\vx_n)\;\;.
\label{Fproperty}
\eq
This establishes the properties of our ensemble of point sets $X$ on which,
in our view, the \qmc\ estimates ought to be based. 

\subsubsection{Estimators}
\label{estimators}
We shall indicate the
`\qmc' nature of the estimators by the superscript $(q)$. The first
estimator is that of the integral:
\bq
E_1^{(q)} = {1\over N}\sum f_j\;\;.
\eq
Here, and in the rest of this section, the sums will run from 1 to $N$.
Denoting by the subscript $(q)$ averages with respect to the `quasi-random'
ensemble discussed above, we then have
\bq
\avgq{E_1^{(q)}} = \intl_C\;f(\vx)\;P_1(s;\vx)\;d^d\vx = J_1\;\;,
\eq
as before: owing to the fact that the one-point distribution is uniform,
the \qmc\ estimate is indeed as unbiased  as the \mc\ one. The
distinction between the two methods appears in the first-order error
estimate. Let us define
\bq
\al(\vx_i,\vx_j) = 1 + F_2(s;\vx_i,\vx_j)\;\;;
\eq
then, we have
\bq
\variq{E_1^{(q)}} = {1\over N}\left(
J_2 - \int f_1f_2\al_{12}\right)\;\;\;+\;\;\;
{\cal O}\left({1\over N^2}\right)\;\;.
\eq
where we have adopted the straightforward convention for integrals
\bq
\int f_1f_2\al_{12} = \intl_C\;f(\vx_1)\;f(\vx_2)\;\al(\vx_1,\vx_2)\;
d^d\vx_1\;d^d\vx_2\;\;,
\eq
etcetera.
As before, we shall insouciantly neglect terms that are sub-leading in $1/N$.
The advantage of the \qmc\ method is now clear: if we can ensure
that $\al_{12}>1$ `where it counts', that is, generally, when $\vx_1$
and $\vx_2$ are `close' in some sense, then the \qmc\ error will be smaller
than the \mc\ one. A good \qmc\ point set, therefore, is one in which the
points `repel' each other to some extent.

The first-order error estimate is now simply
\bq
E_2^{(q)} = {1\over N^2}\sum f_i^2 -
{1\over N^3}\suml f_if_j\al_{ij}\;\;.
\label{QmcErrorEstimator}
\eq
It is simple to show that, indeed
\bq
\avgq{E_2^{(q)}} = \variq{E_1^{(q)}} + {\cal O}(N^{-2})\;\;;
\eq
however, {\em evaluating\/} $E_2^{(q)}$ is less trivial since it is not
obvious how to do this in time linear in $N$. We shall discuss this later.
The variance of the estimator $E_2^{(q)}$ can be evaluated to
\bqa
\vari{E_2^{(q)}} &=& {1\over N^3}\left(
\int f_i^4  - 4\int f_i^3f_j\al_{ij} - \int f_i^2f_j^2\al_{ij}
\right.\nl &&
+ 4\int f_i^2f_kf_l\al_{ik}\al_{kl}
+ 4\int f_i^2f_kf_l\al_{ik}\al_{il}\nl &&\left.
- 4\int f_if_jf_kf_l\al_{ij}\al_{jk}\al_{kl}\right)
 + {\cal O}(N^{-4})\;\;,
\eqa
for which the corresponding estimator (to leading order) is
\bqa
E_4^{(q)} &=& {1\over N^7}\left(
N^3\sum f_i^4  - 4N^2\sum f_i^3f_j\al_{ij} - N^2\sum f_i^2f_j^2\al_{ij}
\right.\nl &&
+ 4N\sum f_i^2f_kf_l\al_{ik}\al_{kl}
+ 4N\sum f_i^2f_kf_l\al_{ik}\al_{il}\nl &&\left.
- 4\sum f_if_jf_kf_l\al_{ij}\al_{jk}\al_{kl}\right)\;\;.
\label{erroronerror}
\eqa
The details of this computation are discussed in the Appendix.
It goes without saying that the substitution $\al_{ij}\to1$ will reduce
all the \qmc\ results to the regular \mc\ ones.\\

We can now see why the `classical' estimator Eq.(\ref{classicalestimator}) overestimates the error. Under
the quasi distribution $P_2$ of Eq.(\ref{multipointdistribution}) the classical estimator averages to 
\bqa
\avgq{E_2}&=&\avgq{{1 \over N^2}\sum_i f_i^2 - {1 \over N^3} \sum_{i,j}f_if_j}\nl 
&= &{1 \over N} (J_2-J_1^2) - {1 \over N^2}\int{f(x)f(y)F(x,y)} +{\cal O}(\frac{1}{N^2})
\eqa
The term involving the correlator is suppressed by ${1 \over N}$, which shows that $E_2$ averages
to something different than the variance of $E_1$ under the quasi distribution. Moreover, we will show 
in section \ref{Laplace} that\footnote{under fairly general conditions for the function $f(x)$.} 
the integral of the suppressed term is strictly positive for
any point-set that is better than a truly random one. So $E_2$ omits a strictly negative term when estimating
the error.\\ 

While it is true that the estimator Eq.(\ref{QmcErrorEstimator}) averages
to a quantity whose leading order in $N$ is equal to the leading order of 
$\vari{E_2^{(q)}}$, it suffers from the following disagreeable property: for 
a constant integrand, while the first two terms vanish identically, the third approaches
zero asymptotically from negative values. This leads to a negative squared error
for all practical purposes. Although this is not disastrous per se, it indicates the 
reason for the appearance of negative errors also for non-constant integrands, as will 
become apparent once we have a concrete expression for the correlation function. It is, thus, 
desirable to obtain an estimator that vanishes identically for constant functions. 
This is achieved by 
\bq
E_2^{(q_2)} = {1 \over \fall{N}{2}}\sum_i f_i^2 -{1\over N\fall{N}{2}}\hat{\sum_{i,j}} f_if_j -{1 \over
N\fall{N}{4}}\hat{\sum_{i,j,k,l}}f_if_j(F_{i,j}-F_{i,k}-F_{l,j}+F_{l,k})
\label{QmcErrorEstimator2}
\eq 
where the $\hat{\Sigma}_{i,j\ldots}$ denotes a sum with all indices different, and $F_{i,j}\equiv F_2(s;\vx_i,\vx_j)$. This quantity averages
to 
\bq
\avg{E_2^{(q_2)}}={1\over N} (J_2-J_1^2)-{1 \over N}\int dxdydzdw\;f(x)f(y)\left[F_{x,y}-F_{x,w}-F_{z,y}+F_{z,w}\right]
\eq
which equals the leading part of $\vari{E_2^{(q)}}$ thanks to Eq.(\ref{vanishing_property}).
It is easy to check that the estimator of Eq.(\ref{QmcErrorEstimator2}) vanishes identically for a constant integrand and any $N$, thanks
to the antisymmetry property of the quadruple sum. 

\subsubsection{Cumulants of $E_1$}
As a final remark, we may also investigate the cumulants of the \qmc\
 estimator $E_1$. We write the expansion of the correlation function
 $F_k$ over $1/N$ as
 \bq
 F_k(s;\vx_1,\ldots,\vx_k)\equiv F_k^{(1)}+\frac{1}{N}F_k^{(2)}+\frac{1}{N^2}F_k^{(3)}+\ldots
 \eq
 and define
 \bq
 \mathcal{M}_{i_1,\ldots,i_k}^{(a)}\equiv \int f(\vx_{1})^{i_1}  \ldots f(\vx_k)^{i_k}F_{k}^{(a)}(s;\vx_1,\ldots,\vx_k)
 \eq
It is evident that if Eq.(\ref{Fproperty}) holds, we have 
\bq
\mathcal{M}_{1,1,\ldots,1}^{(1)} =\frac{ \fall{k}{2}}{2} J_1^{k-2}\mathcal{M}_{1,1}^{(1)}
\eq

The cumulants are defined as
 \bq
 c_n=\avg{\left(E_1^{(q)}-\avg{E_1^{(q)}}_{(q)}\right)^n}_{(q)}
 \eq
 The variance of $E_1$ is then
 \bq
 c_2=\frac{1}{N}(J_2-J_1^2-\mathcal{M}_{1,1}^{(1)}) +O(\frac{1}{N^2})
 \eq
 The skewness is
 \bq
 c_3=\frac{1}{N^2}(J_3-3J_1J_2+2J_1^3-3\mathcal{M}_{1,2}^{(1)}+3J_1\mathcal{M}_{1,1}^{(2)}+6J_1\mathcal{M}_{1,1}^{(1)}-\mathcal{M}_{1,1,1}^{(2)})+O(\frac{1}{N^3})
 \eq
 The unnormalized kurtosis is
\bq
c_4-3c_2^2=\frac{1}{N^2}(-\mathcal{M}_{1,1,1,1}^{(2)}-3(\mathcal{M}_{1,1}^{(1)})^2+4J_1\mathcal{M}_{1,1,1}^{(2)}-6J_1^2\mathcal{M}_{1,1}^{(2)})+O(\frac{1}{N^3})
\eq
The above results indicate that a correlation function that satisfies the property of Eq.(\ref{Fproperty}) leads 
to a distribution whose skewness decreases faster with $N$ than does the variance,
but when it comes to the kurtosis (and higher cumulants), additional 
properties regarding the next-to-leading
order expression for $F$ (denoted above by $\mathcal{M}_{i_1,\ldots,i_k}^{(2)}$) are needed to secure Gaussian 
cumulants\footnote{Approach to a Gaussian distribution,for iid
random variables, would require $c_n/(c_2)^{n/2}$ to approach $0$ for large $N$.}. These properties hold whenever
the saddle point approximation of Eq.(\ref{eerste}-\ref{tweede}) is valid. In such cases one expects Gaussian confidence levels 
for the \qmc\ estimator $E_1$.


\section{Multi-point distributions with diaphonies}

\subsection{Diaphony}
We consider a point set $X$ with $N$ elements, given in $C$.
The non-uniformity of the point set $X$ can be described by its {\em diaphony}\footnote{
some of the concepts of this section have also been discussed 
in \cite{Hoogland} and \cite{Hoogland3}.}:
\bq
D(X) = {1\over N}\suml_{j,k=1}^N\be(\vx_j,\vx_k)\;\;,
\eq
with
\bqa
\be(\vx_j,\vx_k) &=& \nsum\;\sivn\;\en(\vx_j)\ben(\vx_k)\;\;,\nl
\en(\vx) &=& \exp(2i\pi\;\vn\cdot\vx)\;\;.
\eqa
Here, the vectors $\vn=(n_1,n_2,\ldots,n_d)$ form the integer lattice,
and the hat denotes the sum over all $\vn$ except $\vn=\vec{0}$.
We may also write
\bq
D(X) = {1\over N}\nsum\sivn\;\left|\suml_{j=1}^N \en(\vx_j)\right|^2\;\;,
\label{diaphony_def}
\eq
so that we recognize the diaphony as a measure of how well the various
Fourier modes are integrated by the point set $X$. The diaphony is therefore
seen to be related to the `spectral test', well-known in the field 
of random-number generator testing.
For the {\em mode strengths\/} $\sivn$ we have
\bq
\sivn\ge0\;\;\;,\;\;\;\nsum\sivn = 1\;\;.
\label{sigma}
\eq
The latter convention simply establishes the overall normalization of $D$.
The advantage of this diaphony over, say, the usual (star)discrepancy is
the fact that it is translation-invariant:
\bq
\be(\vx_j,\vx_k) = \be(\vx_j-\vx_k)\;\;,
\eq
so that point sets $X$ and $X'$ that differ only by a translation (modulo 1)
have the same non-uniformity: the diaphony is actually defined on the
hyper-torus rather than on the hypercube. Also, the diaphony is
{\em tadpole-free}:
\bq
\intl_C\;\be(\vx)\;d^d\vx = 0\;\;.
\eq
Moreover, we shall use $\sivn$ such that $\sivn=\si^2_{\vn'}$ if the
two lattice vectors $\vn$ and $\vn'$ differ only by a permutation of their
components. Thus, $X$ and $X'$ will also have the same non-uniformity   
if they differ by a global permutation of the coordinates of the points.

\subsubsection{Some numerical results}\label{diaphony_plots}
In this section the behavior of a specific diaphony is presented, for three point sequences, as 
the number of points $N$ increases.

The diaphony is defined by Eq.(\ref{diaphony_def}) with 
\bq
\sigma_{\vn}=Ke^{-\lambda \vn^2}\;\;\;\;K^{-1}=\sum e^{-\lambda \vn^2}\;\;\;\lambda=0.1
\eq
 The reason for experimenting with this definition lies in the factorizing property of the $\sigma_{\vn}$. 
Due to $K^{-1}$ being related to Jacobi theta functions, we call this the `Jacobi diaphony'. We will be using this diaphony
in most of what follows.

In this paper we will be using three point sequences that we will be calling {\tt Ranlux}, {\tt Van Der Corput}
and {\tt Niederreiter}. {\tt Ranlux} is a pseudo-random point sequence generated by
the {\tt Ranlux} algorithm (see \cite{Ranlux}) with luxury level equal to $3$. {\tt Van Der Corput} is a quasi-random sequence
generated by an implementation of the algorithm by Halton that generalizes to many dimensions an older algorithm by
Van der Corput (see \cite{vdcorput}) with prime bases $2,3,5,7,11,\ldots$. Finally {\tt Niederreiter} is another, 
optimal\footnote{in a sense described in \cite{Niederreiter} and \cite{Bratley}.}
quasi-random sequence based on the algorithm in (see \cite{Niederreiter}). In particular, we follow the choices of \cite{Bratley} 
and construct the sequence in whichever base is optimal for the current dimension (see \cite{Niederreiter}).

It should also be noted that only modes with square length up to $\vn^2 \leq 15$ are included 
in the calculation of the diaphony (including the determination of $K$), in anticipation of the same restriction on the estimator
sums in later sections.

In the plots that follow, the diaphony of the {\tt Niederreiter} sequence in particular, but also that of the {\tt van der Corput}
sequence, exhibited a large variation in relatively small intervals of $N$. As the number of points $N$ approaches certain
critical values the diaphony reaches very small levels, only to return to its `cruising' values a few points later. To avoid
cluttering the plots we present here the diaphony averaged in packs of $500$ points without information on the minimum 
or maximum value found in each pack. The minimum values for each pack, that correspond to exceptional point configurations, 
are very interesting on their own but do not affect the present study.
  

\begin{figure}[htbp]
\begin{minipage}[t]{\linewidth}
\begin{tabular}[t]{ll}
\epsfig{file=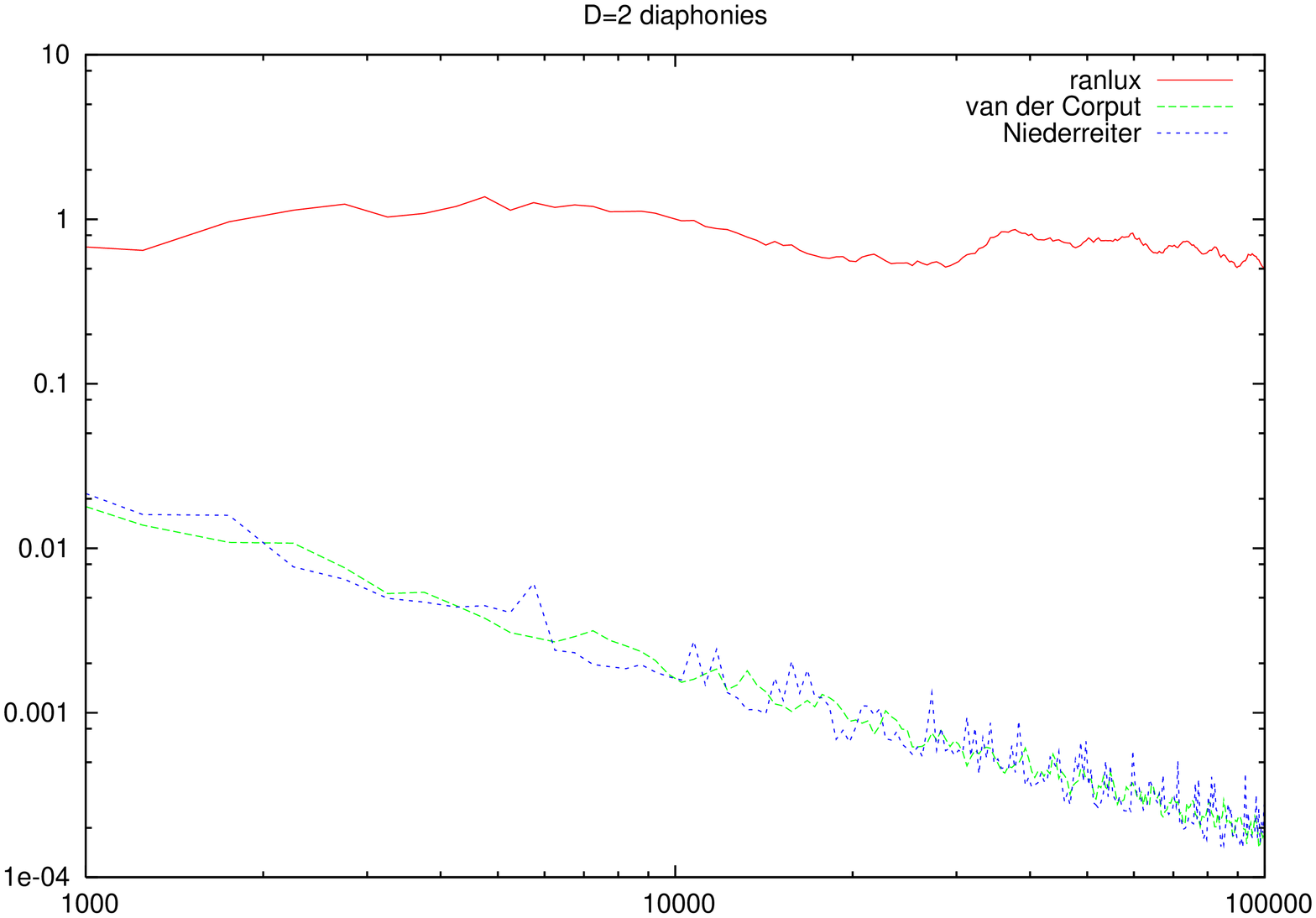,angle=270,width=7cm}
&
\epsfig{file=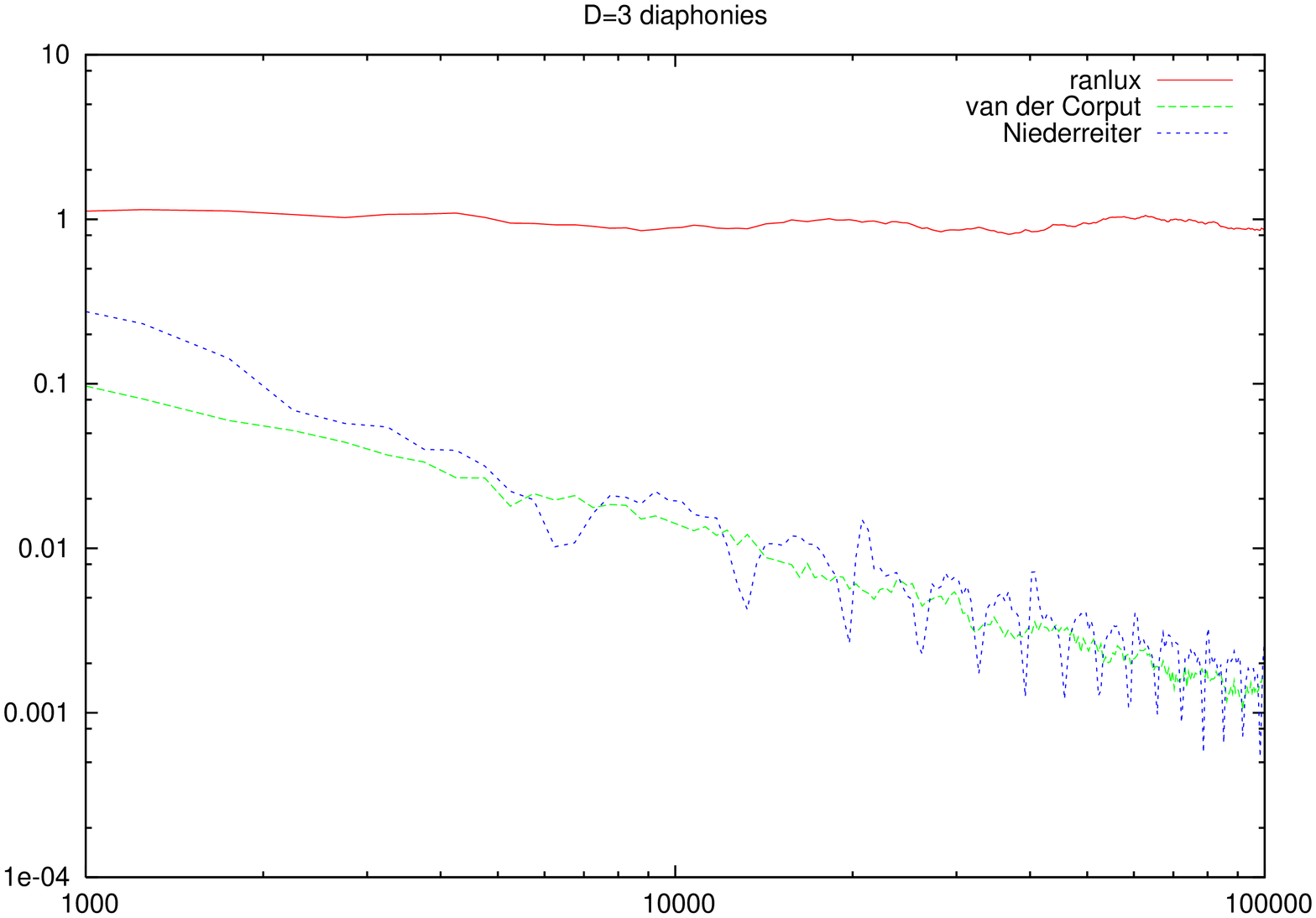,angle=270,width=7cm}
\end{tabular}
\caption{
D=2 (left) and D=3 (right). The diaphony of {\tt RANLUX} (red line), {\tt Van Der Corput} (green line) and {\tt Niederreiter} (blue line).
}
\end{minipage}
\end{figure}

\begin{figure}[htbp]
\begin{minipage}[t]{\linewidth}
\begin{tabular}[t]{ll}
\epsfig{file=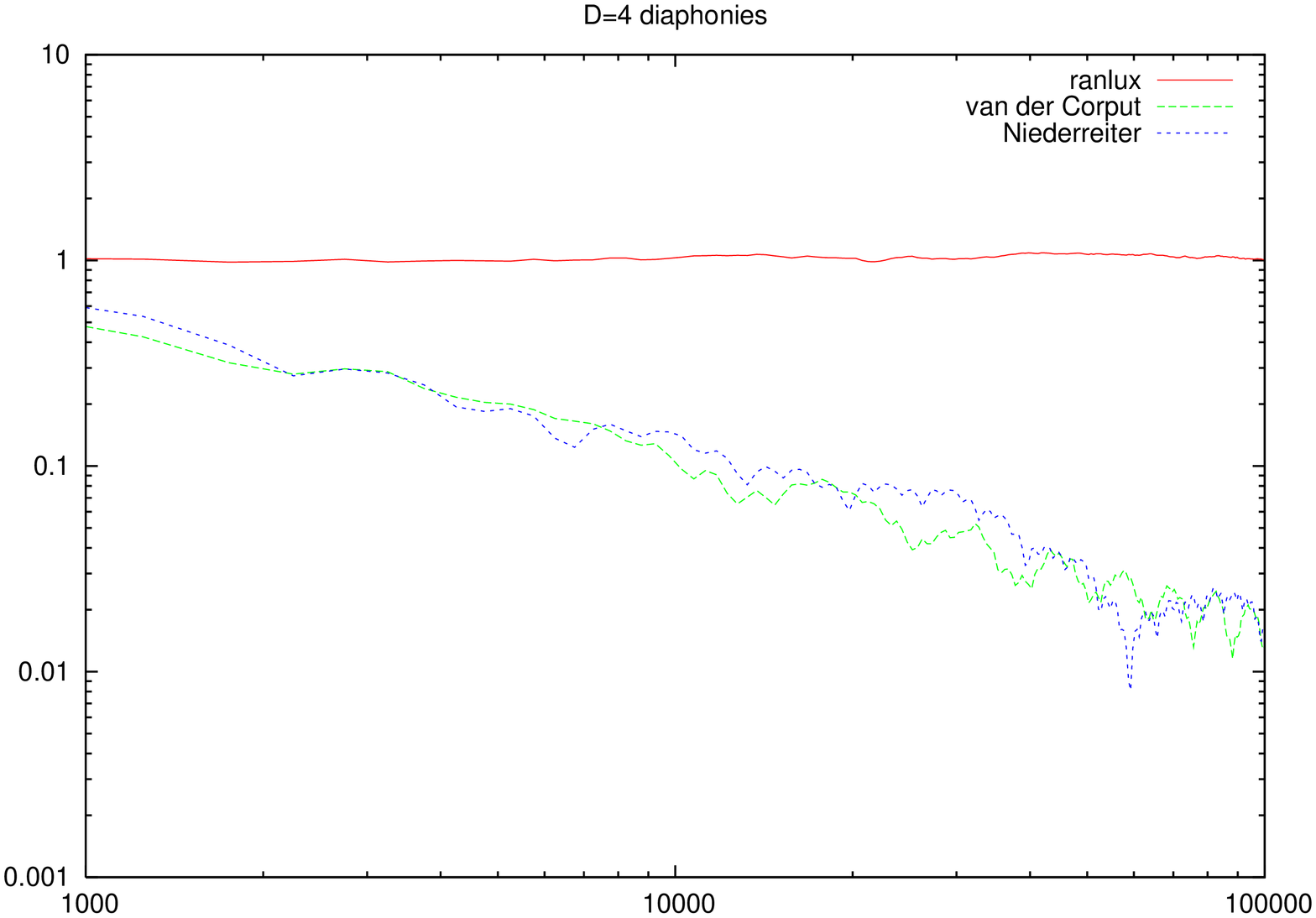,angle=270,width=7cm}
&
\epsfig{file=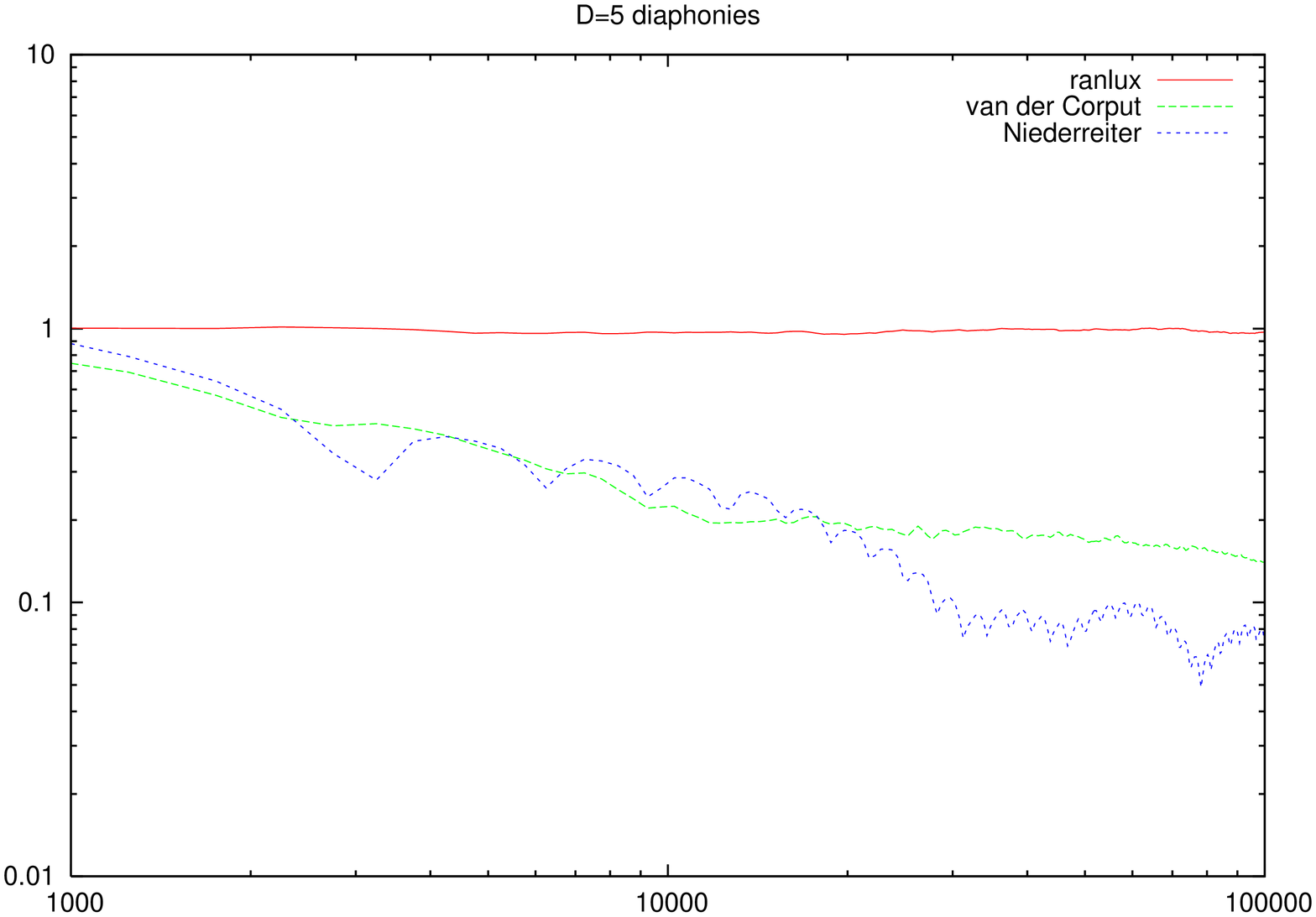,angle=270,width=7cm}
\end{tabular}
\caption{
D=4 (left) and D=5 (right). The diaphony of {\tt RANLUX} (red line), {\tt Van Der Corput} (green line) and {\tt Niederreiter} (blue line).
}
\end{minipage}
\end{figure}


The diaphony of the {\tt RANLUX} sequence is seen to oscillate around 1, as expected.
Moreover the behavior of the {\tt Niederreiter} sequence improves with the number
of dimensions when compared with crude {\tt Van der Corput}, an encouraging
hint for higher dimensions. 
\subsection{Generating function}
We shall now compute a $1/N$ approximation to the moment-generating distribution
of the $p$-point probability distribution, that is,
\bq
G_p(z) = \left\langle\exp(zD(X))\vphantom{A^A}
\right\rangle_{\vx_{p+1},\vx_{p+2},\ldots,\vx_N}\;\;,
\eq
where we have indicated that the points $\vx_1,\vx_2,\ldots,\vx_p$ are kept
fixed while the remaining $N-p$ points are integrated over. $G_p(z)$ therefore
still depends on $\vx_1,\ldots,\vx_p$. This is most easily achieved using
a diagrammatic approach, which has been introduced in \cite{Hoogland}. We shall indicate with crosses those points that are
kept fixed (with an implied sum over them, from 1 to $p$), 
and with dots (`beads') those points that are integrated over 
(again, with an implied sum running from $p+1$ to $N$). The function $\be$
is indicated by a solid line. As the simplest examples, then, we have
\bq\mbox{if $p=N$:}\;\;\;
{1\over N}\plaat{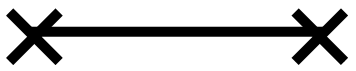}{1.5}{-0.05} = {1\over N}\suml_{j,k=1}^N \be(\vx_j-\vx_k)
=D(X)\;\;,
\eq
and
\bq\mbox{if $p=0$:}\;\;\;
\langle D(X)\rangle_{\vx_1,\ldots,\vx_N} = \be(0) = 
\plaat{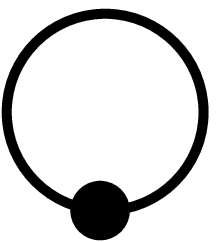}{0.6}{-0.3} = 1\;\;.
\eq
Other examples are
\bqa
\plaat{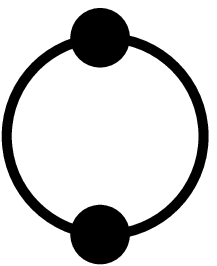}{0.6}{-0.2} &=& 
\intl_C\;\be(\vx_1-\vx_2)^2\;d^d\vx_1\;d^d\vx_2\;\;,\nl
\plaat{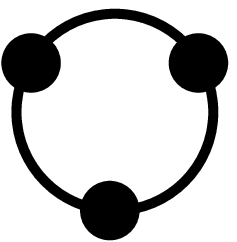}{0.6}{-0.2} &=& 
\intl_C\;\be(\vx_1-\vx_2)\be(\vx_2-\vx_3)\be(\vx_3-\vx_1)\;
d^d\vx_1\;d^d\vx_2\;d^d\vx_3\;\;,
\eqa
and so on: a general closed loop with precisely $n$ beads will be denoted
by $\plaat{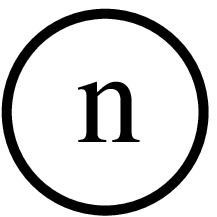}{0.5}{-0.15}$. Note that, since, the functions $\en(\vx)$
form an orthonormal (and even a complete) set, we have
\bq
\plaat{fig05}{0.7}{-0.3} = \nsum\left(\sivn\right)^n\;\;.
\eq
We can now simply write out all possible (connected and disconnected) diagrams
where every solid line ends in a cross or a bead, and apply the following
Feynman rules:
\begin{enumerate}
\item A factor $2z/N$ for every $\be$ line (where the factor 2 arises from the
two possible orientations);
\item A factor $\fall{(N-p)}{q}$ for every diagram (or product of diagrams)
that contains precisely $q$ beads\footnote{The `falling power' is defined
as $\fall{a}{b} = a!/(a-b)! = a(a-1)(a-2)\cdots(a-b+1)$.};
\item In addition, the usual symmetry factors arising from equivalent
lines and vertices, and from the repetition of identical (sub)diagrams.
\end{enumerate}
We shall compute $G_p(z)$ including terms of order 1 and those of order $1/N$.
Note that 
\bq
\fall{(N-p)}{q} = N^q\left(1-{pq\over N}-{q(q-1)\over2N}\right)
+ {\cal O}(N^{-2})
\eq
as long as $N\gg pq,q^2$. In the following we shall always
assume this.\\

First, we consider contributions without any crosses or nontrivial vertices.
A general term in this class is given by
\nc{\necklaces}{{1\over r_1!}\left(z\;\plaat{fig01}{0.6}{-0.25}\right)^{r_1}
{1\over r_2!}\left(z^2\;\plaat{fig02}{0.6}{-0.25}\right)^{r_2}
{1\over r_3!}\left({4z^3\over3}\;\plaat{fig03}{0.7}{-0.25}\right)^{r_3}
\cdots\;\;,}
$${\fall{(N-p)}{Q}\over N^Q}\necklaces
$$
where
\bq
Q = r_1 + 2r_2 + 3r_3 + \cdots\;\;;
\eq
up to order $1/N^2$, this contribution to the generating function can 
therefore be written as
\nc{\chains}{\suml_{\{r\}}\prol_{n\ge1}{1\over r_n!}
\left({(2z)^n\over2n}\plaat{fig05}{0.6}{-0.2}\right)^{r_n}}
\bqa
G_p^{(1)}(z) &=& \left(1 - {pz\over N}{\der\over\der z} - 
{z^2\over2N}{\der^2\over\der z^2}\right)\chains\nl
&=&\left(1 - {pz\over N}{\der\over\der z} - 
{z^2\over2N}{\der^2\over\der z^2}\right)G^{(0)}(z)\;\;,\nl
G^{(0)}(z) &=&
\exp\left(-{1\over2}\nsum\log\left(1-2z\sivn\right)\right)\;\;.
\eqa
Up to $1/N^2$, one diagram with a four-point vertex may be present:
a generic contribution of this type is
$$
{\fall{(N-p)}{Q+m+1+m_2+1}\over N^{Q+m_1+m_2+2}}
\left({(2z)^{m_1+m_2+2}\over8}\;\;\plaat{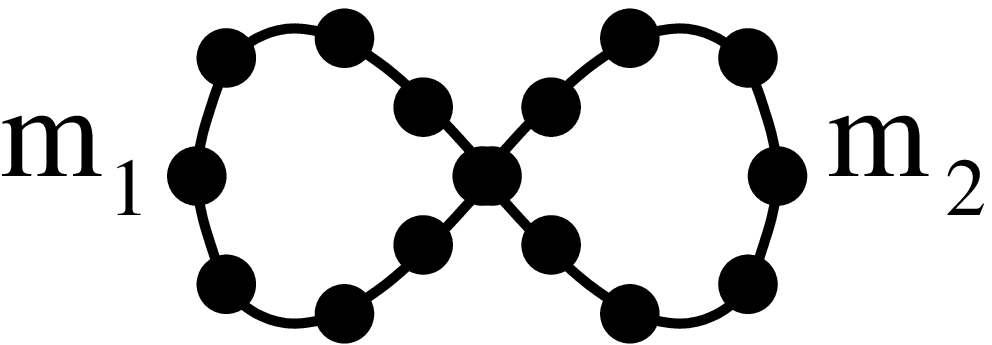}{3}{-0.25}\right)
$$ $$\times\necklaces
$$
where $m_{1,2}$ denote the number of beads on each loop, excluding the
one on the four-vertex. Let us define
\bq
\phi(z;\vx_j-\vx_k) = \nsum
{2z\sivn\over1-2z\sivn}\en(\vx_j)\ben(\vx_k)\;\;;
\eq
then, this contribution can be written as
\bq
G_p^{(2)}(z) = {1\over8N}\phi(z;0)^2G^{(0)}(z)\;\;.
\eq
Note that the lemniscate graph is actually equal to the product of
two closed loops: this is a consequence of the translational invariance
of the diaphony.
A generic contribution containing two three-vertices is 
$$
{\fall{(N-p)}{Q+m_1+m_2+m_3+2}\over N^{Q+m_1+m_2+m_3+3}}
\left({(2z)^{m_1+m_2+m_3+3}\over12}\;\;\plaat{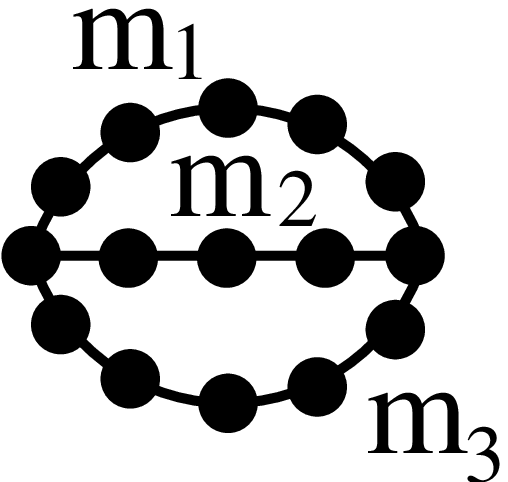}{2}{-0.8}\right)
$$ $$\times\necklaces
$$
so that this contribution to the generating function reads
\bq
G_p^{(3)}(z) = {1\over12N}G^{(0)}(z)\;\intl_C\;\phi(z;\vx)^3\;d^d\vx\;\;.
\eq
The diagrams with crosses have the generic contribution
$$
{\fall{(N-p)}{Q+m}\over N^{Q+m+1}}
\left(z(2z)^{m}\;\;\plaat{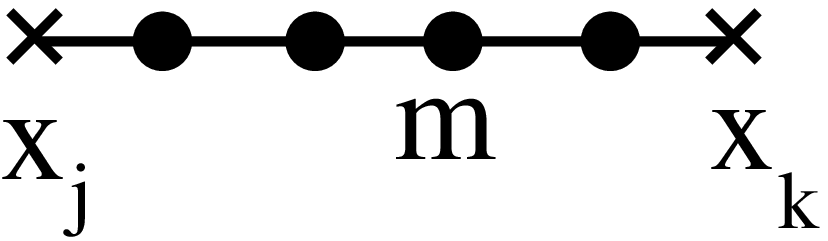}{2.5}{-0.5}\right)
$$ $$\times\necklaces
$$
leading to
\bqa
G_p^{(4)}(z) &=& {1\over2N}G^{(0)}(z)\;\suml_{j,k=1}^p
\phi(z;\vx_j-\vx_k)\nl
&=& {1\over N}G^{(0)}(z)\left(
{p\over2}\phi(z;0)^2 + \suml_{1\le j<k\le p}\phi(z;\vx_j-\vx_k)
\right)\;\;,
\eqa
where we have singled out the contributions with $j=k$.
All other possible diagrams either vanish because of translational
invariance and tadpole-freedom, or are of order $1/N^2$ or lower.
The final result for the generating function
up to order $1/N^2$ is therefore
\bqa
G_p(z) &=& G^{(0)}(z)\left( 1
- {1\over 4N}\intl_C\;\phi(z;\vx)^2\;d^d\vx
+ {1\over 12N}\intl_C\;\phi(z;\vx)^3\;d^d\vx
\right.\nl &&\left.\hspace{2cm}
+\;{1\over N}\suml_{1\le j<k\le p}\phi(z;\vx_j-\vx_k)
\vphantom{\intl_C}\right)\;\;.\label{generatingfunction}
\eqa
Note that the term in $G^{(1)}(z)$ containing $p$ cancels precisely against
that in $G^{(4)}(z)$, so that the only reference to $p$ is in the
last term in brackets in Eq.(\ref{generatingfunction}), and indeed we have
\bq
\intl_C\;G_p(z)\;d^d\vx_p = G_{p-1}(z)\;\;.
\label{recursive}
\eq

In Appendix B we give the result for the higher order ($O(\frac{1}{N^2})$) term in $G_p$. 
There are 25 terms that contribute but only three of them include $p$. The condition \ref{recursive}
still holds. 


\subsection{Multi-point distribution by Laplace transform}
\label{Laplace}
From the generating function, we can recover the actual probability
distributions. As discussed above, 
let $H(s)$ be the probability that the point set
$X$ has diaphony equal to $s$, that is, $D(X)=s$. The underlying ensemble
of point sets is that of sets of $N$ iid uniformly distributed points,
{\it i.e.\/} the same ensemble underlying the usual Monte Carlo 
error estimates. Then, we have
\bqa
H(s) &=& \intl_C\;d^d\vx_1\;d^d\vx_2\cdots d^d\vx_N\;
\delta(D(X)-s)\nl
&=& {1\over2i\pi}\intii\;e^{-zs}\;G_0(z)\;dz\;\;,
\eqa
where the integration contour runs to the left of all the singularities of 
$G_0(z)$; and the multi-point distribution for $p$ points averaged over all
point sets $X$ with diaphony $s$, is given by
\bqa
P_p(s;\vx_1,\vx_2,\ldots,\vx_p) &=& {1\over H(s)}R_p(s;\vx_1,\vx_2,\ldots,\vx_p)
\;\;,\nl
R_p(s;\vx_1,\vx_2,\ldots,\vx_p) &=& {1\over2i\pi}\intii\;
e^{-zs}\;G_p(z)\;dz\;\;.
\eqa
Since we write the deviation from uniformity of the multi-point distribution as
\bq
P_p(s;\vx_1,\vx_2,\ldots,\vx_p) = 
1 - {1\over N}F_p(s;\vx_1,\vx_2,\ldots,\vx_p)\;\;,
\eq
we see that the multi-point correlation $F_p$ is, up to $O(\frac{1}{N})$, as claimed,
built up from two-point correlators\footnote{This doesn't hold for the next order in $\frac{1}{N}$
as seen in Appendix B. Terms like the one of Eq.(\ref{ThreePointFunction}), that don't
factorize, appear for $p\ge3$.}: for $p\ge3$,
\bq
F_p(s;\vx_1,\vx_2,\ldots,\vx_p) = F_{p-1}(s;\vx_1,\vx_2,\ldots,\vx_{p-1})
+ \suml_{j=1}^{p-1}F_2(s;\vx_j,\vx_p)\;\;,
\eq
so that the $p$-point correlator is simply the sum of all $p(p-1)/2$
2-point correlators.
In the approximation used, the sub-leading terms in $H(s)$ are
actually irrelevant, and we may write
\bqa\hspace{-2cm}
H(s) &\approx& {1\over2i\pi}\intii \exp(\psi(s;z))\;dz\;\;,\nl
\psi(s;z) &=& - sz - {1\over2}\nsum\log\left(1-2z\sivn\right)\;\;,\nl
F_2(s;\vx_1,\vx_2) &=& {-1\over2 \pi i H(s)}\intii
\exp(\psi(s;z))\phi(z;\vx_1-\vx_2)\;dz\;\;.\label{fullformulae}
\eqa
Except in the very simplest cases\footnote{See section \ref{box_approximation}.}, a complete evaluation of
Eq.(\ref{fullformulae}) is nontrivial. A simplification arises if
$s$ is much smaller than its expectation value 1 (which is anyway the aim
in quasi-Monte Carlo), or if the Gaussian limit
is applicable, namely when the number of modes with
non-negligible $\sivn$ becomes large in such a way that no single
mode dominates. In practice, this happens when the dimensionality of $C$
becomes large. Fortunately, these are precisely the situations of interest.
The position of the saddle point for $H(s)$, $\hz$, is given by
\bq
\nsum{\sivn\over1-2\hz\sivn} = s\;\;.\label{eerste}
\eq
For $s\ll 1$, therefore, $\hz$ is large and negative. Since to first order
the same saddle point may be used for $R_2$, we find the attractive
result
\bq
F_2(s;\vx_1,\vx_2) \approx \nsum\ovn\;
\en(\vx_1)\;\ben(\vx_2)\;\;\;,\;\;\;
\ovn = {2\hz\sivn\over2\hz\sivn-1}\;\;.\label{tweede}
\eq
The formulae (\ref{eerste}) and (\ref{tweede}) suffice, in our approximation,
to compute all the multi-point correlations.\\

We finish this section with the following observation. Suppose that $F_2$
is given as a function of $\vx_1,\vx_2$. By Fourier integration we 
can then compute the $\ovn$. The assumption that the
saddle-point approximation is valid, together with the
normalization condition $\sum\sivn=1$, then allows us to write
\bq
\hz = -{1\over2}\nsum{\ovn\over1-\ovn}\;\;\;,\;\;\;
\sivn = -{1\over2\hz}{\ovn\over1-\ovn}\;\;\;,\;\;\;
s = \nsum\sivn(1-\ovn)\;\;.
\eq
We see that $F_2$ not only determines the {\em form\/} of the diaphony, but in
addition also its {\em value}. 


\section{Application of \qmc\ estimators}

\subsection{The mechanism behind error reduction}
After the above preliminaries we can now examine the mechanism by which
\qmc\ can outdo \mc. 
We shall assume the saddle-point approximation to be valid.
For $s<1$, we then have $\hz<0$, and all the $\ovn$ are positive, and
as $\hz\to-\infty$ they approach unity from below (although for
$|\vn|\to\infty$ they must always, of course, go to zero). Now notice that the set
of functions $\en(\vx)$ is complete, that is,
\bq
\suml_{\vn}\;\en(\vx_1)\;\ben(\vx_2) = \delta^d(\vx_1-\vx_2)\;\;.
\eq
This allows us to write the variance of the \mc\ error as
\bq
\vari{E_1} = {1\over N}\;\nsum\;
\left|\;\intl_C\;f(\vx)\;\en(\vx)\;d^d\vx\;\right|^2\;\;,
\eq
where the contribution from the zero mode $\vn=0$ is canceled by the
$J_1^2$ term. For \qmc\, on the other hand, we find
\bq
\vari{E_1^{(q)}}_{(q)} = {1\over N}\;\nsum\;(1-\ovn)
\left|\;\intl_C\;f(\vx)\;\en(\vx)\;d^d\vx\;\right|^2\;\;.    
\label{variance}
\eq
We see that those modes $\vn$ for which $\ovn$ is positive
tend to lead to an error reduction. In the saddle-point approximation,
therefore, {\em any\/} value $0<s<1$ will lead to a decreased error
with respect to standard \mc. On the other hand, since 
\bq
0<\hz<\min_{\vn}{1\over2\sivn}\;\;\;\mbox{for}\;\;s>1\;\;,
\eq
large values of the diaphony will actually lead to an {\em increase\/}
in the error.
Note that in the above we have only used the fact that the $\en$ form
a {\em complete, orthonormal\/} set of functions: therefore, the error-reduction
result holds for a much wider class of discrepancies than just the diaphonies
discussed in this paper.

\subsection{Estimators analyzed}
We can now arrive at an estimator for the \qmc\ error.
The simplest form is obtained by inserting Eq.(\ref{tweede}) in the equation for $E_2$ (Eq.\ref{QmcErrorEstimator}):
\bq
E^{(q)}_2=\frac{1}{N^2}\sum f_i^2 - \frac{1}{N^3}(\sum f_i)^2 - \frac{1}{N^3}\sum_{\hat{\vn}}\omega_{\vn} | \sum_i f_i e_{\vn}(x_i)|^2
\label{estimatorfinal}
\eq
with
\bq
\omega_{\vn}=\frac{-2\hat{z} \sigma_{\vn}^2}{1-2\hat{z} \sigma_{\vn}^2}
\eq

We are still free to choose the exact form of the weights $\sigma_{\vn}^2$ at will, under the constraints
of Eq.(\ref{sigma}). Our choice is the so called Jacobi weights\footnote{due to their convenient factorizing property.}
\bq
\sigma_{\vn}^2=K e^{-\lambda \vn^2}
\eq
with
\bq
K^{-1}=\hat{\sum}_{\vn}e^{-\lambda \vn^2}
\eq                     

The parameter $\lambda$ controls the `sensitivity' of the diaphony: as $\lambda \rightarrow 0 $ we get $ \sigma_{\vn} \rightarrow 1$
for every mode which corresponds to a super-sensitive diaphony, useless for practical purposes,
while as $\lambda \rightarrow \infty$ only the modes with $\vn^2=1$ contribute making the diaphony fairly non-sensitive.
We choose $\lambda=0.1$. Other values of $\lambda$, 
within a `reasonable range' do not alter, in practice, the numerical value of $E_2^{(q)}$, as shown in section \ref{raising_lambda}.

It is easy to see  that the estimator averages (to leading order in $N$) in a positive definite quantity 
\footnote{It averages to Eq.(\ref{variance}) which is positive definite as long as $s<1$.}. This leaves still open the
possibility for a negative error estimate, particularly for relatively smooth functions where
the cancellation between the two sums of the pseudo estimate are large leading to a small error.
The source of the negative error effect is clear in the case of a constant function. Then
\bq
f(x)=C\Rightarrow E_2^{(q)}=-{1 \over N^3} C^2 \sum_{\vn}\omega_{\vn}\sum_{i,j}u_{\vn}(x_i)\bar{u}_{\vn}(x_j)
\eq
and the point sum of every Fourier mode can be anything from $0$ (when the points are spread evenly enough to 
produce complete cancellations for all the included modes)
to $N^2$ (when all the points are on top of each other). The average of this sum is $N$ (for truly random points), but
for \qmc\ points we expect that this sum will be significantly smaller than that. For non-constant functions similar effects
can be expected, apart from the fact that the first two terms of $E_2^{(q)}$ do not cancel anymore. Thus, we expect negative
squared errors for higher modes or small number of points, and this is what has been observed in a number of plots. Unfortunately
there is no way to predict precisely when, as $N$ increases, the estimator gets a useful, positive value. One could resort to
the error of $E_2^{(q)}$, but that is cubic in the number of modes (see Eq.\ref{erroronerror}) and hence prohibitively expensive 
in realistic calculations.\\

The way out of this is the estimator of Eq.(\ref{QmcErrorEstimator2}) which can be written in a form with unrestricted sums as follows:
\bqa
E_2^{(q)}&=&{1 \over \fall{N}{2}}S_2-{1 \over N\fall{N}{2}}S_1^2-\frac{\fall{(N-1)}{3}}{N\fall{N}{4}}\sum_{\vn}\omega_{\vn}|W_{\vn}|^2 \nl
	 & &+\frac{\fall{(N-1)}{3}}{N\fall{N}{4}}S_2\sum_{\vn}\omega_{\vn}
	 +\frac{N-1}{N\fall{N}{4}}\sum_{\vn}\omega_n(2S_1\Re\left\{W_{\vn}\bar{U}_{\vn}\right\}-2\Re\left\{U_{\vn}\bar{Q}_{\vn}\right\})\nl
	 & &-\frac{1}{N\fall{N}{4}}\sum_{\vn}\omega_{\vn}(N-2+|U_{\vn}|^2)(S_2-S_1^2)
\label{final2}
\eqa
where  
\bqa
U_{\vn}\equiv\sum_i u_{\vn}(x_i)\nl
W_{\vn}\equiv\sum_i u_{\vn}(x_i) f(x_i)\nl
Q_{\vn}\equiv\sum_i u_{\vn}(x_i) f^2(x_i)\nl
S_2\equiv\sum_i f_i^2 \nl
S_1\equiv\sum_i f_i
\eqa

It is identically zero for a constant function, as can be easily checked, and averages to the leading order of the squared variance of $E_1$.
The correction terms are of higher than leading order in $N$, but that does \emph{not} mean that we have selectively included some NLO corrections
to the variance. The correction terms above are such that the NLO terms vanish on the average.

In practice the infinite sum over modes in both estimators has to be truncated. This should not be perceived as an approximation
of any kind. It amounts to a redefinition of the diaphony.
Looking at Eq.(\ref{eerste}) we see that as the value of $s$ becomes small
the saddle point becomes quickly large and negative: $\hat{z}\ll 0$.
Then $-2\hat{z} \sigma_{\vn}^2\rightarrow \infty$ for low modes and $-2\hat{z} \sigma_{\vn}^2\rightarrow 0$ for higher modes,
when $\sigma_{\vn}^2 / |\hat{z}| \rightarrow 0$. We can, thus,  safely neglect these higher  modes in the estimator. As long as the value
of the diaphony is small, which is in any case the goal in \qmc\  , the profile of $\omega_{\vn}$ depends only on the choice of
$\lambda$, which, as said, also regulates the sensitivity of the diaphony. We see therefore that the estimator inherits the sensitivity of the
diaphony in a direct way.

It is worth noting that the factorized form of the $\beta$-function in the diaphony definition
is directly responsible for the fact that the two estimators are now of complexity $N\times M$ (with $M$ the number
of modes) instead of quadratic in $N$. This is a desirable achievement as long as $M\le N$, which we shall
always assume to be the case.

\subsection{Numerical results}

In the following we will present a number of plots that show how both the `classical' and the quasi error 
estimates\footnote{The `classical' or `pseudo' estimator, $E_2$, is the one of Eq.(\ref{classicalestimator}), constructed
on the assumption that the points are iid. By `quasi' estimator, $E_2^{(q)}$, we mean  the `improved' estimator
Eq.(\ref{final2}).}
behave as a function of the number of points $N$. In the process we will use the three types of point sequences 
defined in section \ref{diaphony_plots}. 

A number of test functions were used for integrands. They
consist of a subset of the test functions used by Schlier in \cite{Schlier}, along with a Gaussian function with
dimension-dependent width. We have
\bq
TF13:f(\vx)=\prod_{k=1}^D \frac{|4x_k-2|+k}{1+k}
\eq
which averages to $J_1=1$. This test function is especially tailored for a Van der Corput sequence,
since in $D=1$ it is perfectly integrated by such a sequence with base $2$.

\bq
TF2:f(\vx)=\prod_{k=1}^D kcos(kx_k)
\eq
which averages to $J_1=\prod_k sin(k)$. This function should be difficult to integrate in high dimensions.

\bq
TF4:f(\vx)=\sum_{k=1}^D\prod_{j=1}^k x_j
\eq
which averages to $J_1=1-\frac{1}{2^D}$. It is chosen as a simple example of a function that is
not a product of single-variable functions.\\

A Gaussian with fixed width suffers from a rapid decrease, in higher dimensions,
of the region of the integration volume where the function is non-zero,
making the integration cumbersome (the higher the dimension,  the more points are needed
and inter-dimensional comparison is difficult). To avoid this we use instead
\bq
TF6:f(\vx)=\prod_{i=1}^D\sum_{n_i=-\infty}^{\infty}\frac{e^{-(x_i-x_{0i}+n_i)^2/2\sigma^2}}{\sqrt{2\pi\sigma^2}}
\eq
which is a product of superpositions of a Gaussian and its tails outside the $[0,1]$ interval. We wish to keep the variance of 
this function independent of the number of dimensions, so we define $\sigma$ such that
\bq
\frac{1}{2\sigma}\sum_{m=-\infty}^{\infty}e^{-m^2/4\sigma^2}=(1+V)^{1/D}\sqrt{\pi}
\eq
where in practice it suffices to keep the first couple of terms in the sum. 
The function averages to $J_1=1$ and spreads as the number of dimension grows ($\sigma\rightarrow\infty$ as $D\rightarrow \infty$).

In the following plots the error and its estimates as functions of the number
of points $N$ are shown in a double logarithmic
scale.

\begin{table}
\begin{center}
\begin{tabular}[t]{lr}

\begin{tabular}[t]{c|c}\hline\hline
D & \# of modes \\ 
\hline
1 & 6\\
2 & 44\\
3 & 250\\
4 & 1256\\
5 & 5182 
\end{tabular}
\qquad\qquad
&
\qquad\qquad
\begin{tabular}[t]{c|c}\hline\hline
D & \# of modes \\ 
\hline
1 & 4\\
2 & 20\\
3 & 56\\
4 & 136\\
5 & 332 

\end{tabular}

\end{tabular}
\caption{number of modes with $\vn^2\leq 15$ (left) and $\vn^2\leq 5$ (right)}
\label{number_of_modes}
\end{center}
\end{table}

The `classical' error estimate is presented,
along with three versions of quasi error estimators, $E^{q5}_2,E^{q10}_2,E^{q15}_2$.
The superscript next to $q$ denotes the squared length of the highest modes included
in the sums of Eq.(\ref{final2}). Thus $E^{q10}_2$ includes\footnote{please
note that  the square length of a mode is the sum of the squares of $D$ 
integers. So 
for $D=2$, for example, the modes present are those with square equal to $1,2,4,5,9,10,13,16,17,\ldots$ and, thus,
$E_2^{(q15)}$ actually contains modes with squared length up to 13.} modes with
$\vn^2 \leq 10$. In table \ref{number_of_modes} we give the number of modes with $\vn^2\leq 15$,
and $\vn\leq 5$ for different dimensions. It is evident that the number of modes grows rapidly with the dimensionality.

 The real error made is included for comparison. 
The data were collected in a point per point basis up to $N=10^5$. In the plots we have included
the average value of each error for successive subsets of $500$ points, suppressing any information
on minimum or maximum values in the subset\footnote{The 
real error (in particular) fluctuates a lot
as the quasi sets complete their successive cycles of low diaphony, but knowledge of the specific
point where the error minimizes is of course not available a priori.}.

All integrations are
performed in the unit hypercube $[0,1]^D$. The dimensionality varies from 2 to 6.

\begin{figure}[htbp]
\begin{minipage}[t]{\linewidth}
\begin{tabular}[t]{ll}
\epsfig{file=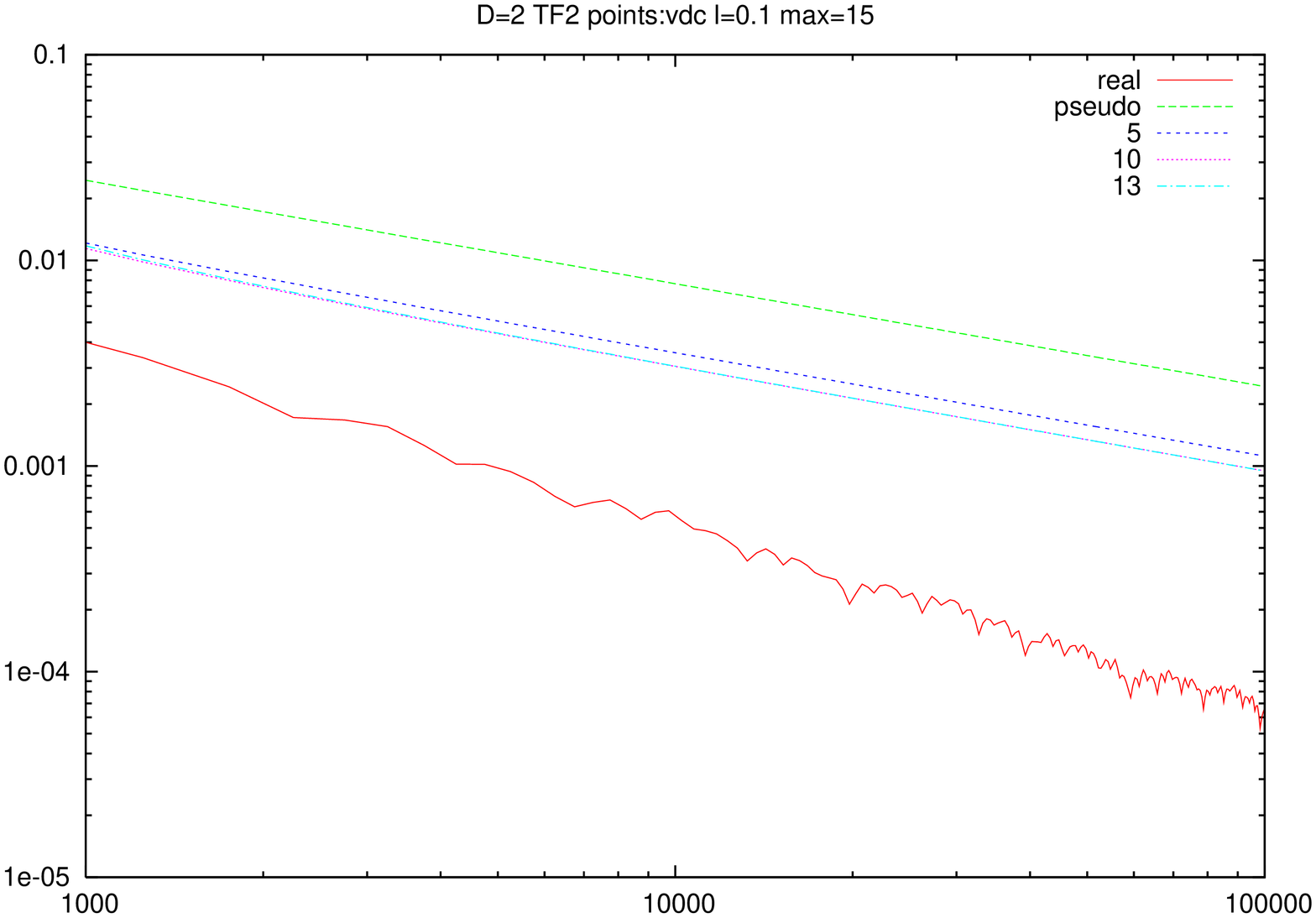,angle=270,width=7cm}
&
\epsfig{file=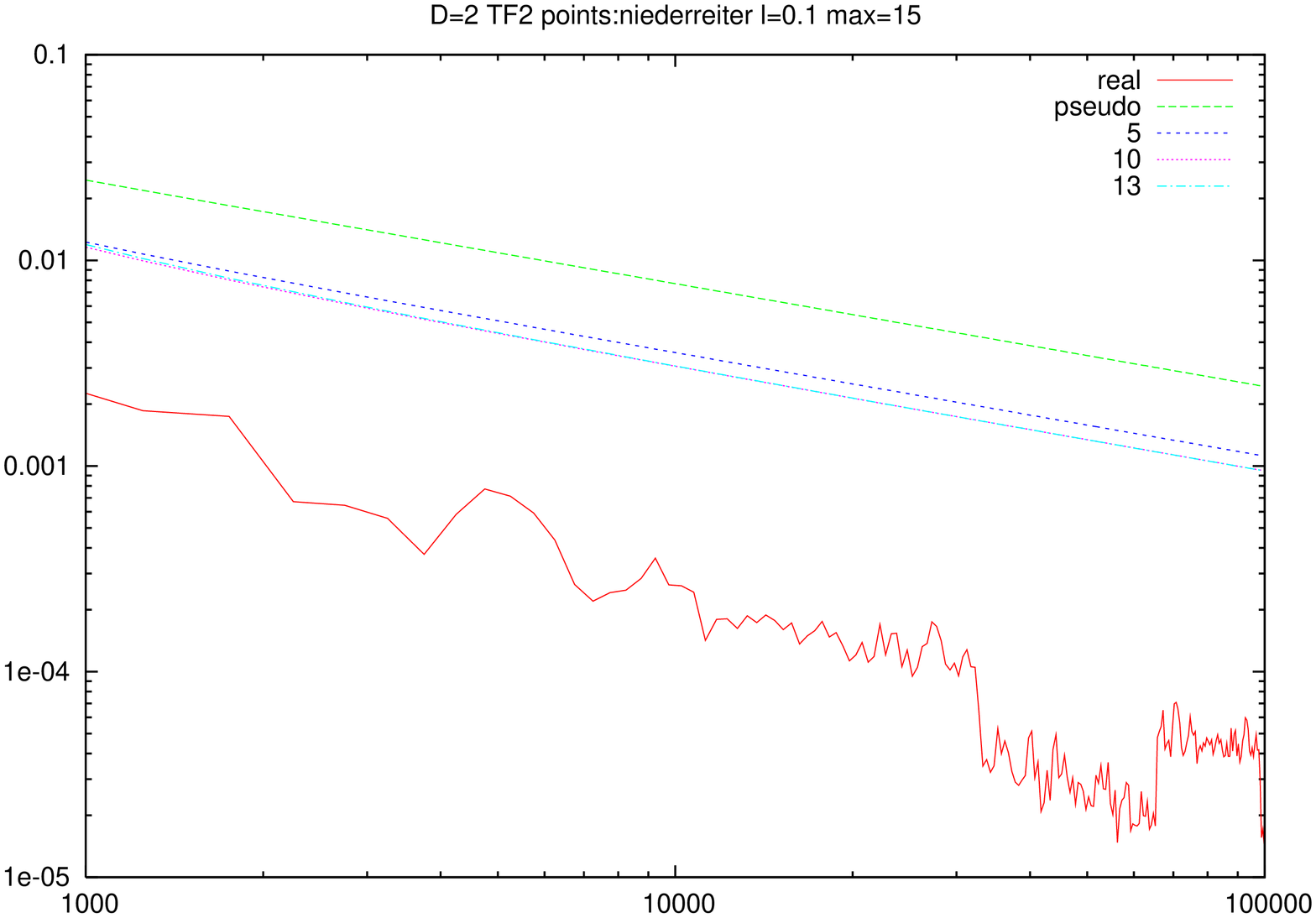,angle=270,width=7cm}
\\
\end{tabular}
\caption[TF2,d=2]{
TF2, d=2 log-plot using a {\tt Van der Corput} sequence (left) and a {\tt Niederreiter} sequence (right). 
The classical error estimator is far off the real error
whereas the quasi estimators are approaching the real error as more modes are added to the sum.
The need for more modes is, however, obvious, in both plots. 
}
\label{TF2d2_vdc}
\end{minipage}
\end{figure}
\begin{figure}[htbp]
\begin{minipage}[t]{\linewidth}
\begin{tabular}[t]{ll}
\epsfig{file=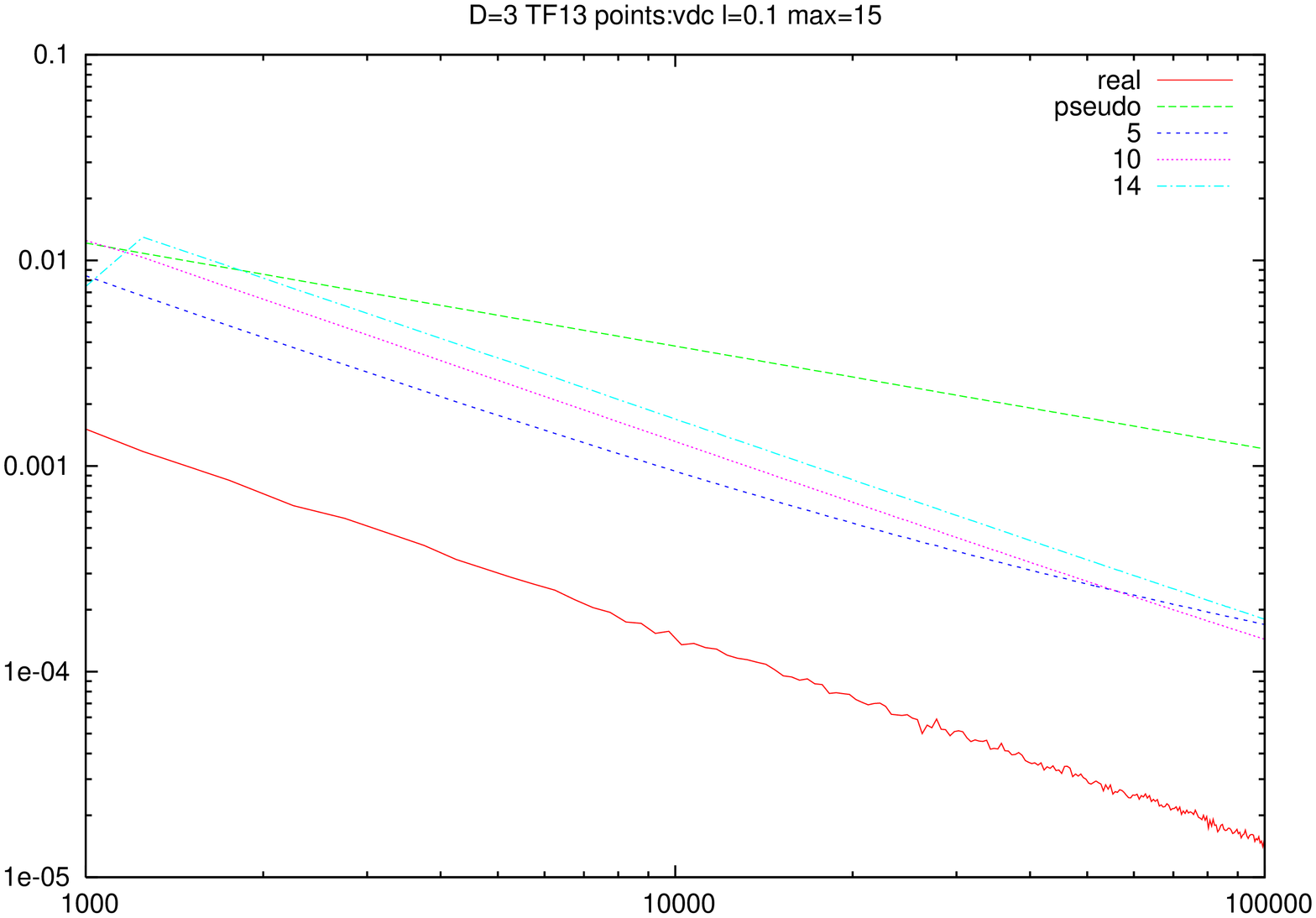,angle=270,width=7cm}
&
\epsfig{file=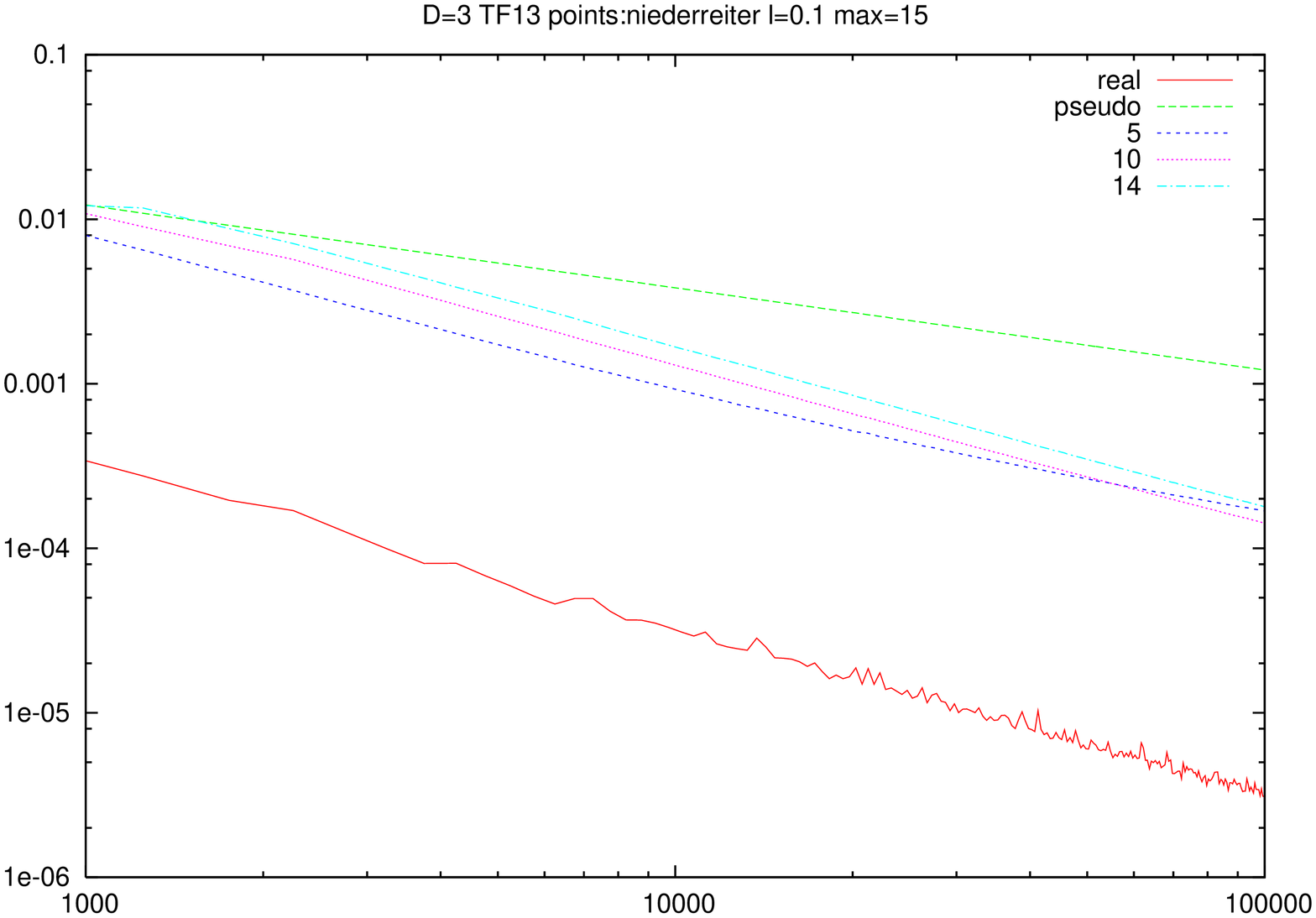,angle=270,width=7cm}
\end{tabular}
\caption[TF13, d=3]{
TF13, d=3 log-plot using a {\tt Van der Corput} sequence (left) and a {\tt Niederreiter} sequence (right). 
The quasi estimators follow the error with the appropriate $N$-dependence contrary to the pseudo estimator.
Note that the $E^{q14}$ is in this case worse than $E^{q10}$ or $E^{q5}$ for all $N\leq 100000$. The higher modes
converge slower to their average value, but the cross-over point is not known in advance and it is function-dependent.
}

\end{minipage}
\end{figure}

\begin{figure}[htbp]
\begin{minipage}[t]{\linewidth}
\begin{tabular}[t]{ll}
\epsfig{file=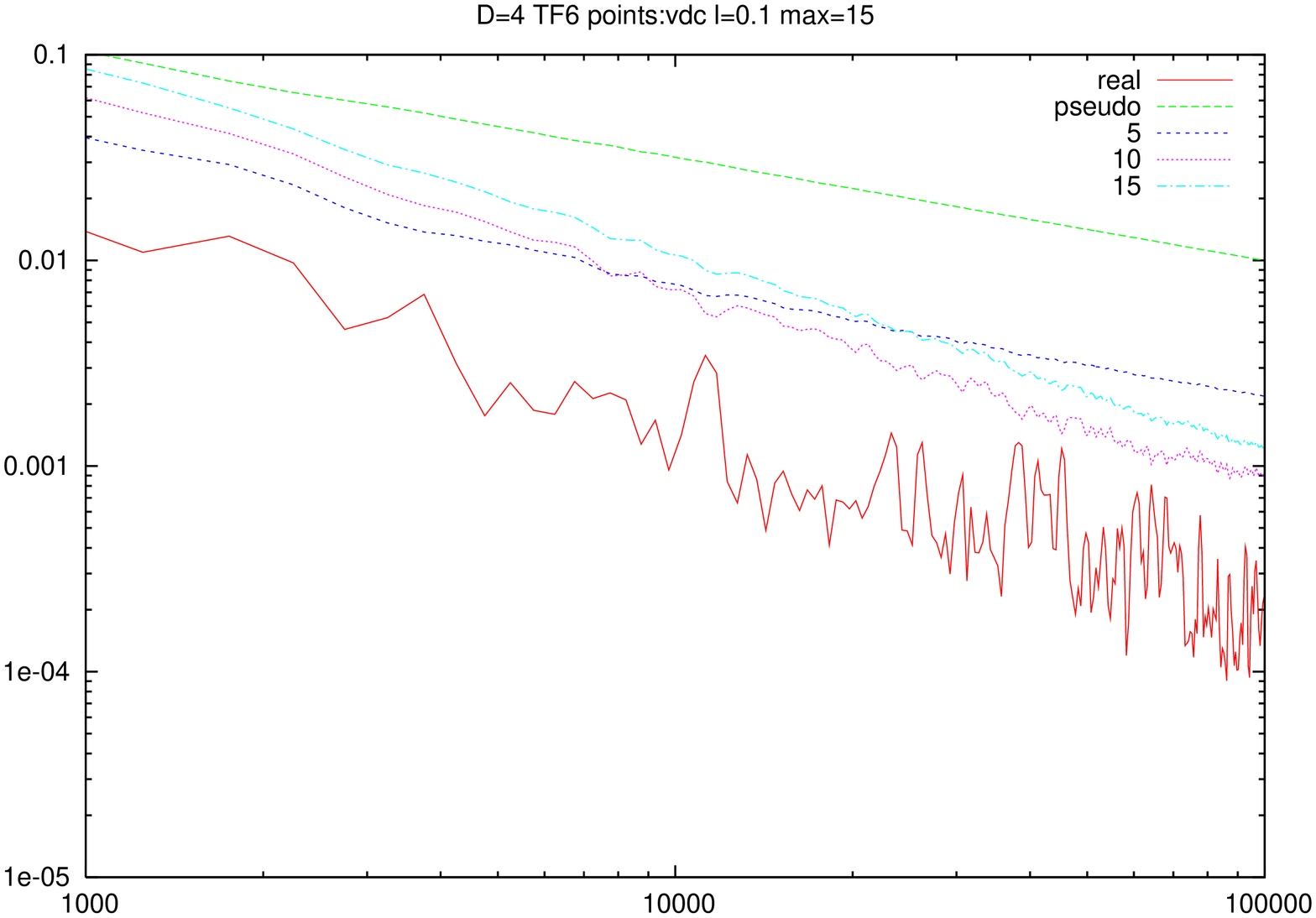, angle=270, width=7cm}
&
\epsfig{file=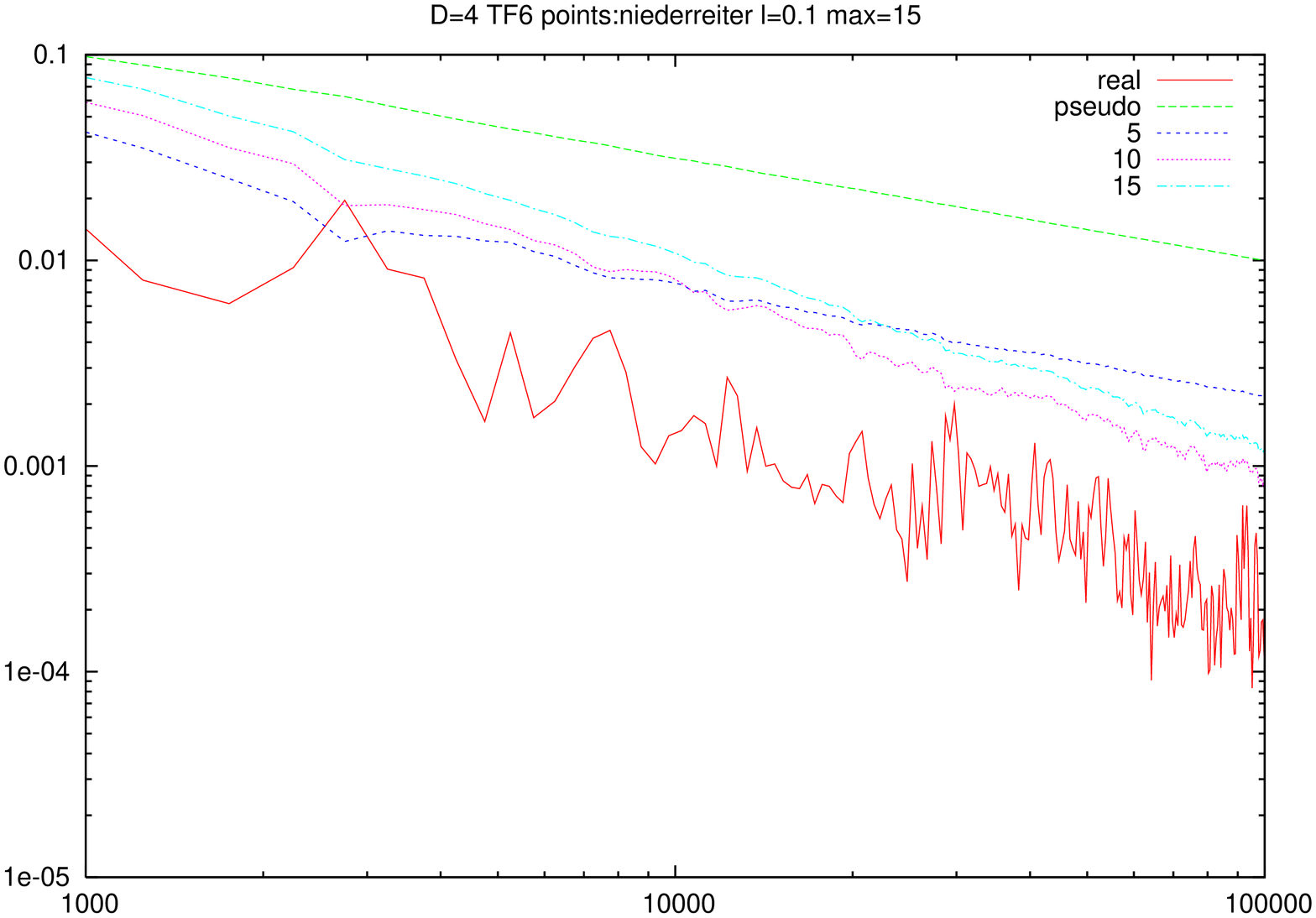, angle=270, width=7cm}
\end{tabular}
\caption[TF6, d=4]
{
TF6, d=4 log-plot using a {\tt Van der Corput} sequence (left) and a {\tt Niederreiter} sequence (right).
The quasi estimators approximate well the error. Moreover we see here a clearer instance of the crossover of 
higher modes in large $N$ mentioned in the previous figure.
}
\label{TF5d4}
\end{minipage}
\end{figure}
\begin{figure}
\begin{minipage}[t]{\linewidth}
\begin{tabular}[t]{ll}
\epsfig{file=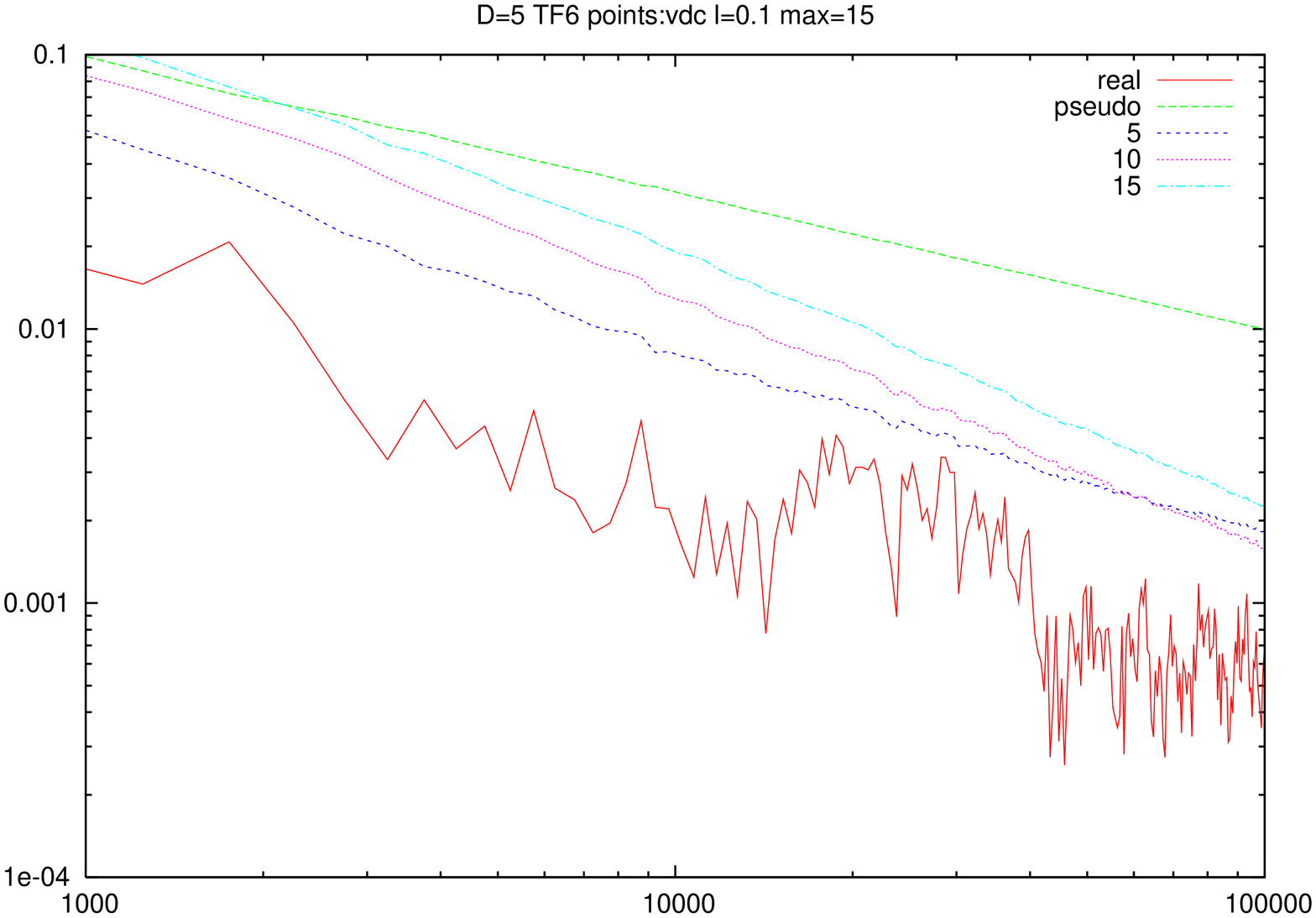, angle=270, width=7cm}&
\epsfig{file=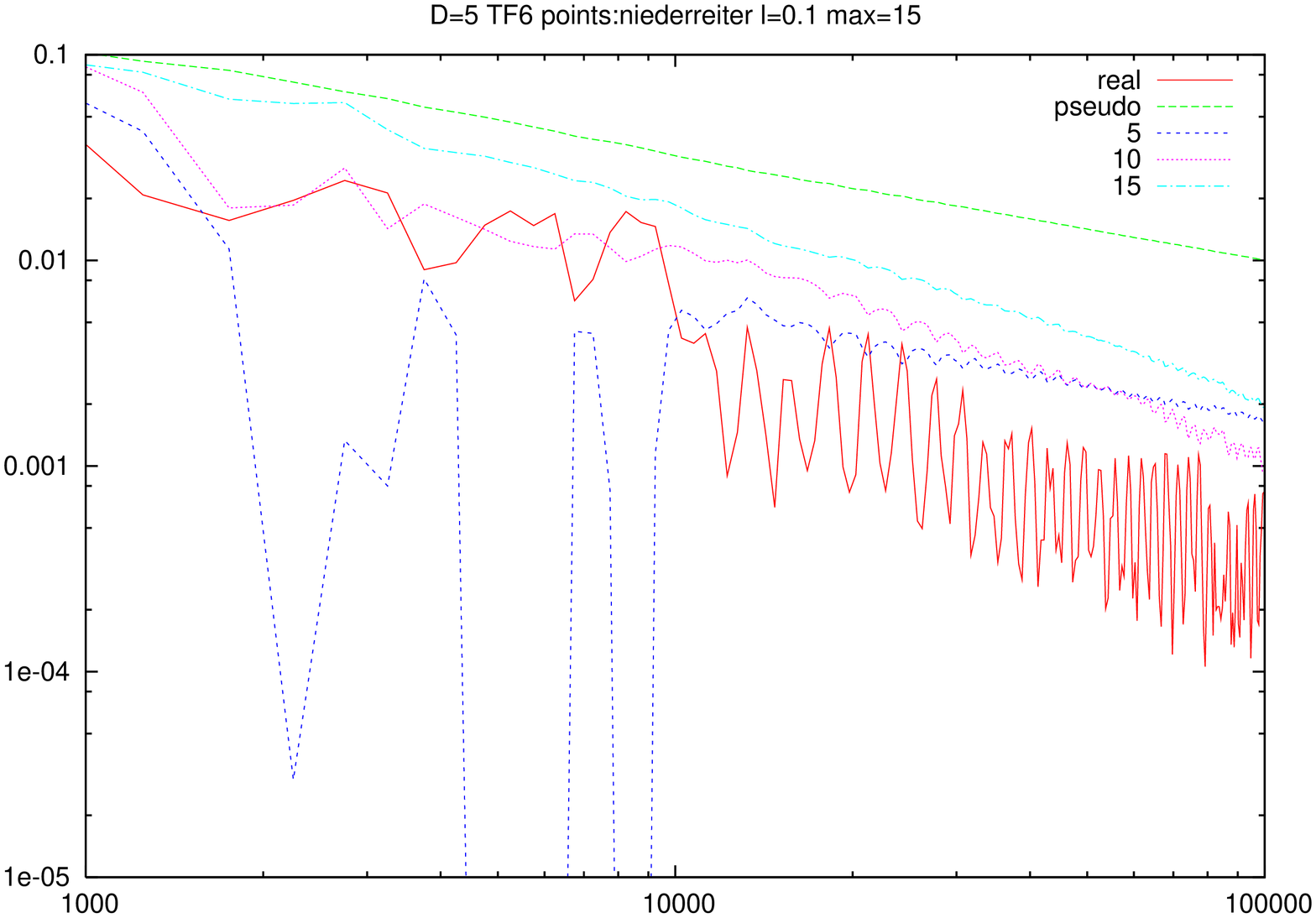, angle=270, width=7cm} 
\end{tabular}
\caption[TF6, d=5]
{
TF6, d=5 log-plot using a {\tt Van der Corput} sequence (left) and a {\tt Niederreiter} sequence (right). 
The use of the improved estimator (Eq.\ref{final2}) reduces the probability of a negative error square
estimate but, naturally, it doesn't remove it altogether. The plot on the right demonstrates this effect. As 
expected, the estimator returns to positive values and stabilizes as the number of points increases and the estimator
converges to its average value. 

}
\end{minipage}
\end{figure}
\begin{figure}
\begin{minipage}[t]{\linewidth}
\begin{tabular}[t]{ll}
\epsfig{file=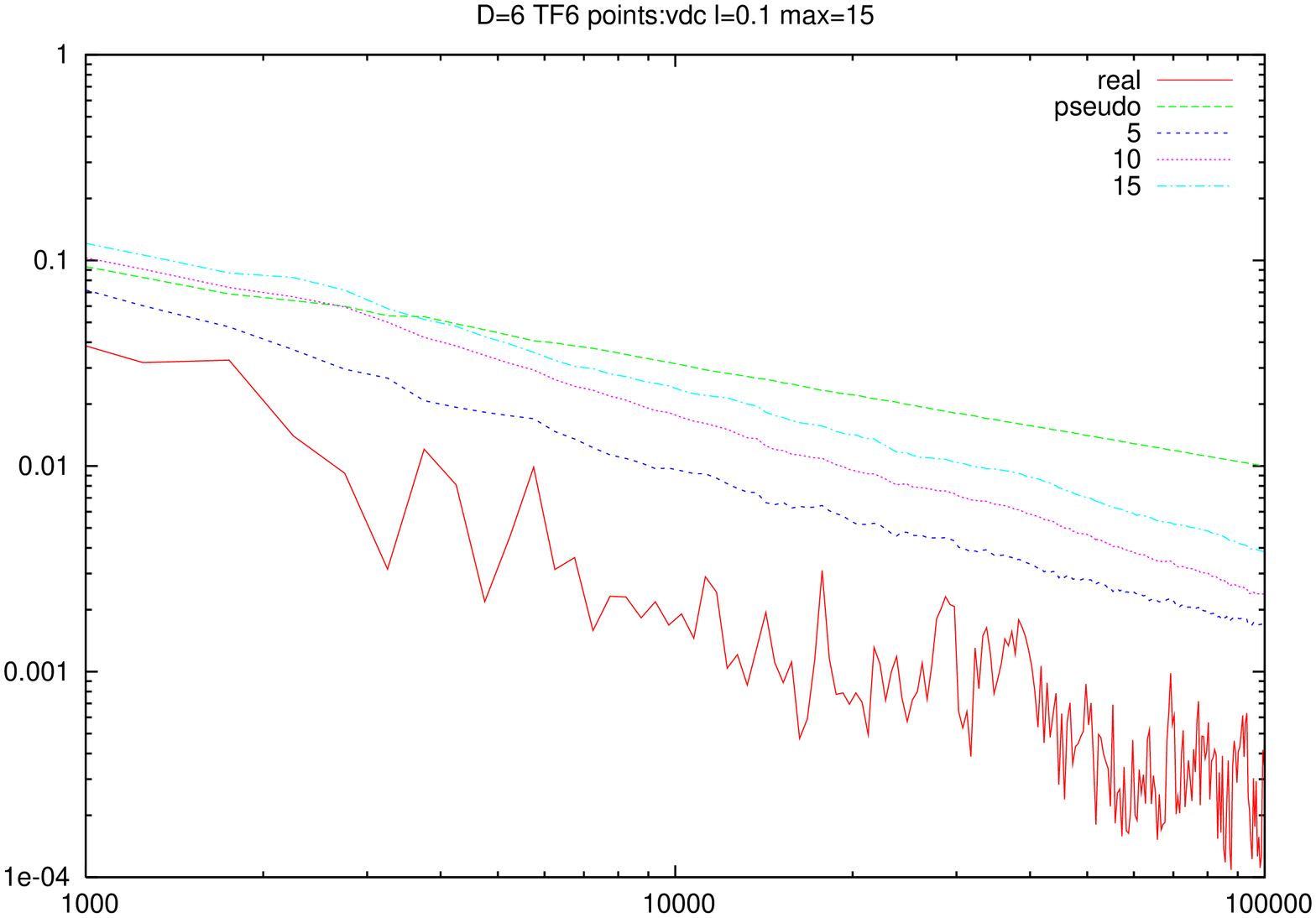, angle=270, width=7cm}&
\epsfig{file=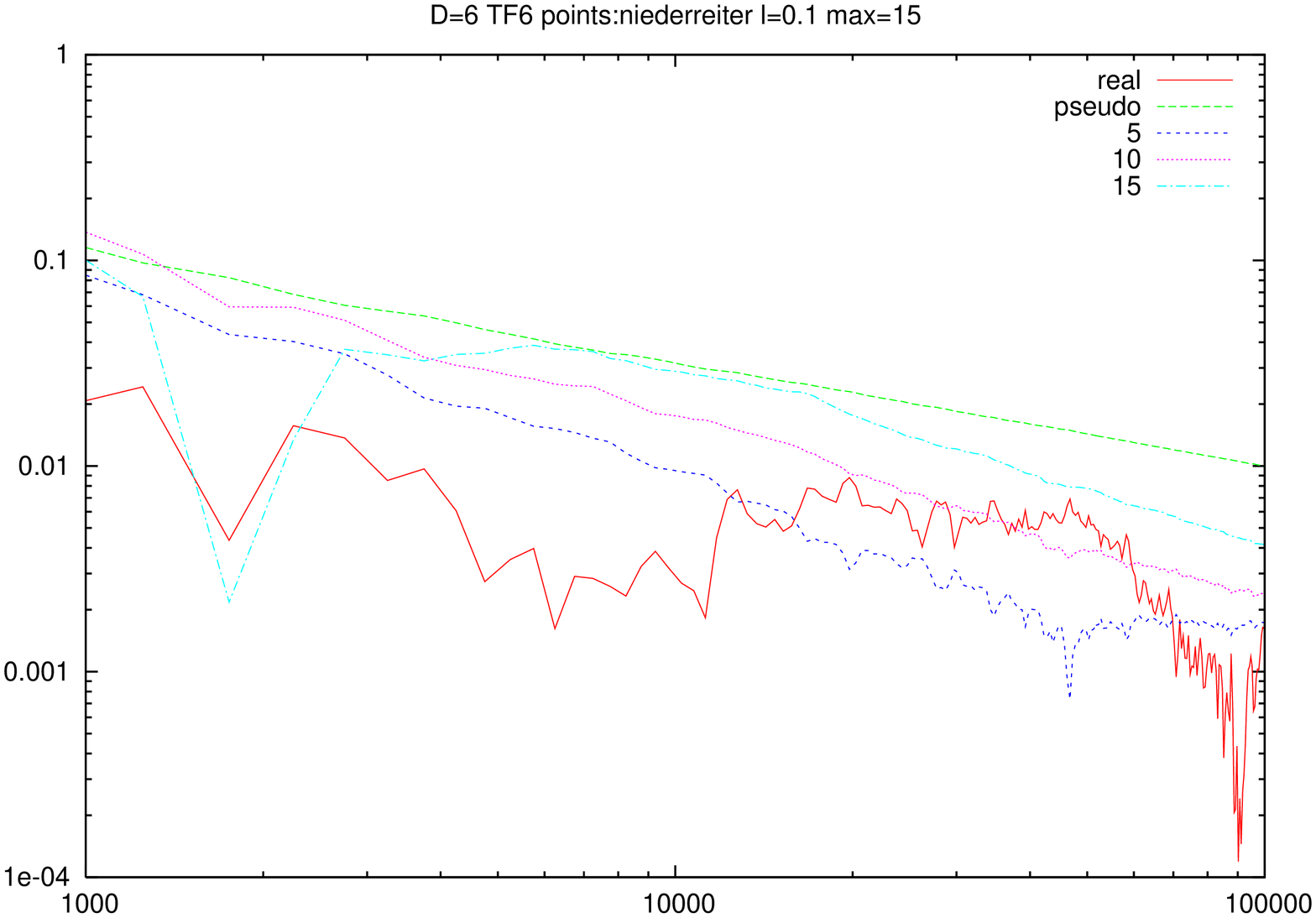, angle=270, width=7cm} 
\end{tabular}
\caption[TF6, d=6]
{
TF6, d=6 log-plot using a {\tt Van der Corput} sequence (left) and a {\tt Niederreiter} sequence (right).
In this case the estimators describe very well the real error made in the integration. 
}
\end{minipage}
\end{figure}

\clearpage


\section{Alternative approaches}

\subsection{Raising the value of $\lambda$ in the Jacobi diaphony}
\label{raising_lambda}
In general the real \qmc\ error is approached by including more and more modes
in the estimator sum. At the same time, by including higher modes, one increases
the error on this estimate (the error on $E_2$) because one attempts to estimate
by Monte Carlo means the integral $\int f(\vx) \en(\vx)$ which will fluctuate vigorously 
for higher modes. 

One might then attempt to raise the value of $\lambda$, 
thus decreasing the number of active
modes (that give an appreciably non-zero $\omega_{\vn})$). This would of 
course reduce
the sensitivity of the diaphony, artificially lowering its value. 
Improvement in the error estimate originating from higher modes 
would be lost but the contribution
of the modes close to the origin (which are the ones included) would 
be relatively enhanced, as can be seen from the behavior of 
the weights $\omega_{\vn}$ (see Eq.\ref{tweede}) . 
\begin{figure}[htbp]
\begin{minipage}[t]{\linewidth}
\begin{tabular}[t]{ll}
\epsfig{file=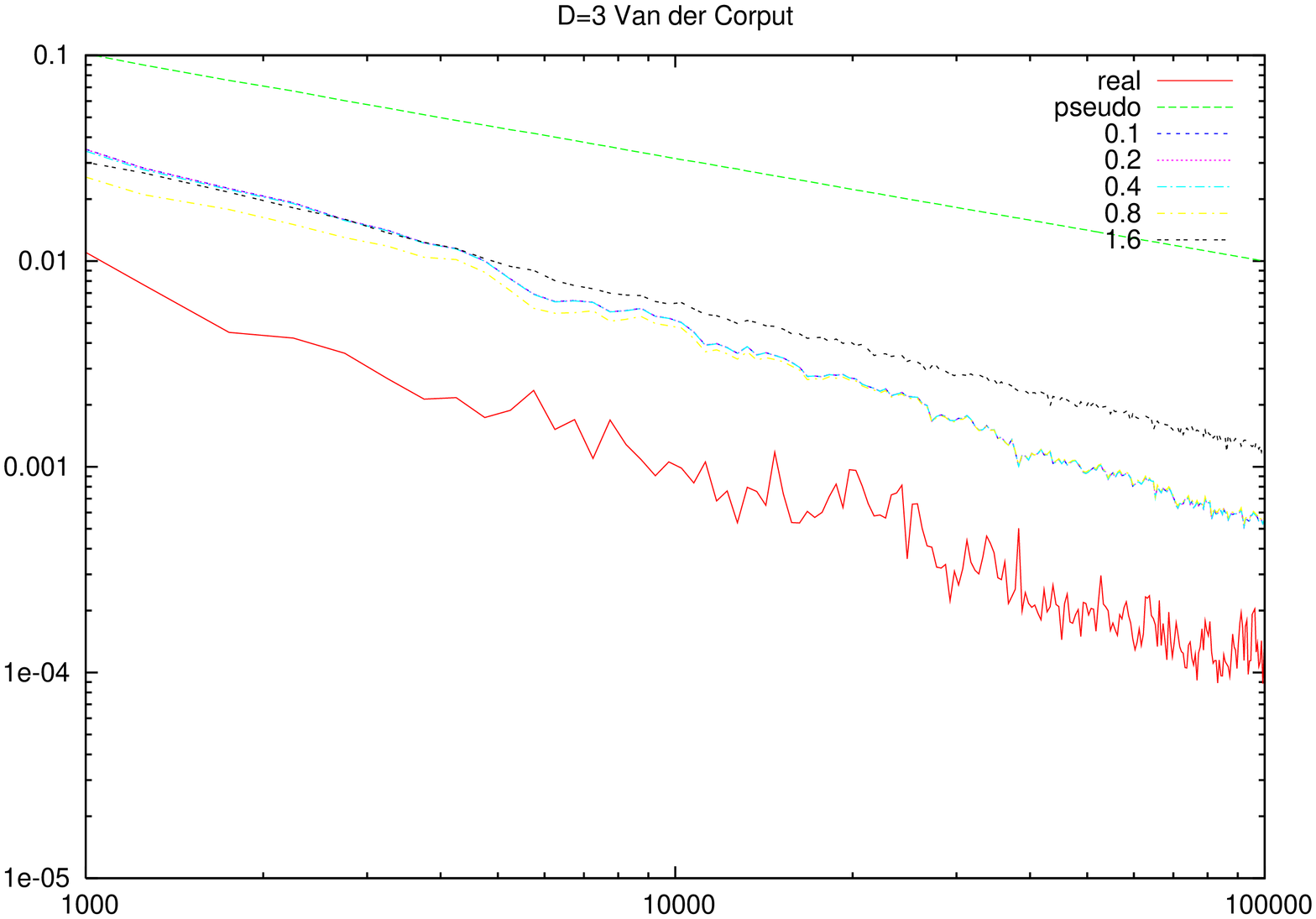, angle=270, width=7cm}
&
\epsfig{file=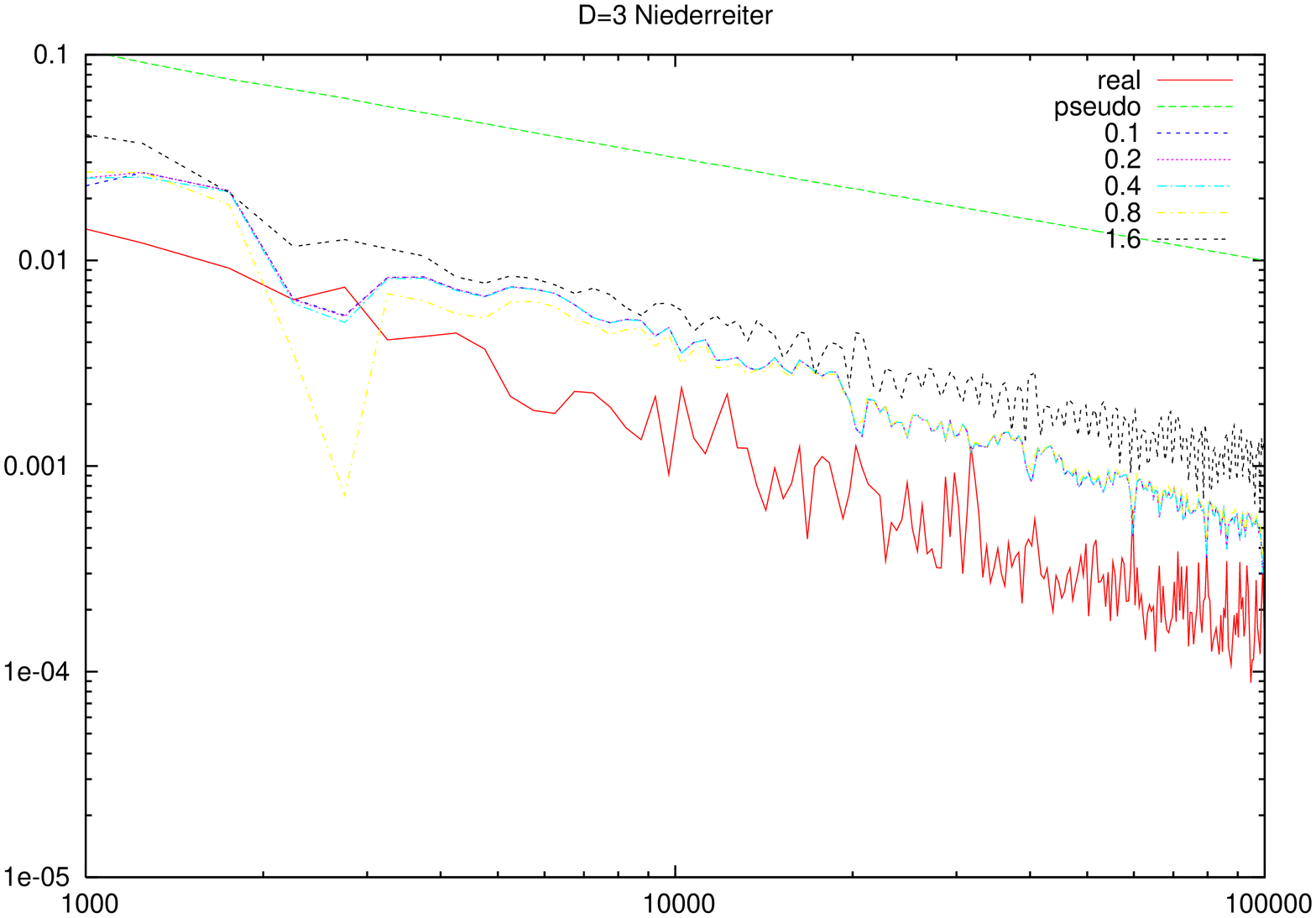, angle=270, width=7cm}
\end{tabular}
\caption[TF6, d=4]{
TF6, d=4 log-plot using the {\tt Van der Corput} (left) and the {\tt Niederreiter} (right) sequences. $E^{q15}$ is shown
for different values of $\lambda$ indicated in the key, along with the real error and the classical estimate.
Average values of all quantities for sets of 500 points are shown in each case.
The value of $\lambda$ doesn't alter the estimator, as long as 
that value stays within a specific range. We see that, in this case, the value $\lambda=1.6$ is out of the safe range.
}
\end{minipage}
\end{figure}
\clearpage


\subsection{Monitored estimator}
\label{monitoring}
The estimators $E^{q5}_2$ , $E^{q10}_2$ and $E^{q15}_2$ are not always proportional
to the classical estimate, and, in some cases they decreases quite faster with $N$ than the classical estimate does.
They never decrease slower than the classical estimate, though, and one can use that as follows.
One monitors the ratio of $E^{q15}_2$, 
for example, to the `classical' error estimate, and after a certain point\footnote{which 
depends on the resources of the user.}, the `classical' error is only
estimated and multiplied with that ratio. This  is a purely linear algorithm and therefore very fast.
Caution has to be exercised, though, in the way the critical ration is chosen, in order to avoid 
configurations where the estimators acquire a very low value for some exceptional value of $N$.

This approach relies heavily on the, frequently false, assumption that the quasi and classical estimators
scale. If this is not so, the new estimate is conservative. One has, thus,
the option to trade accuracy for cpu time.

\begin{figure}
\begin{minipage}[t]{\linewidth}
\begin{tabular}[t]{ll}
\epsfig{file=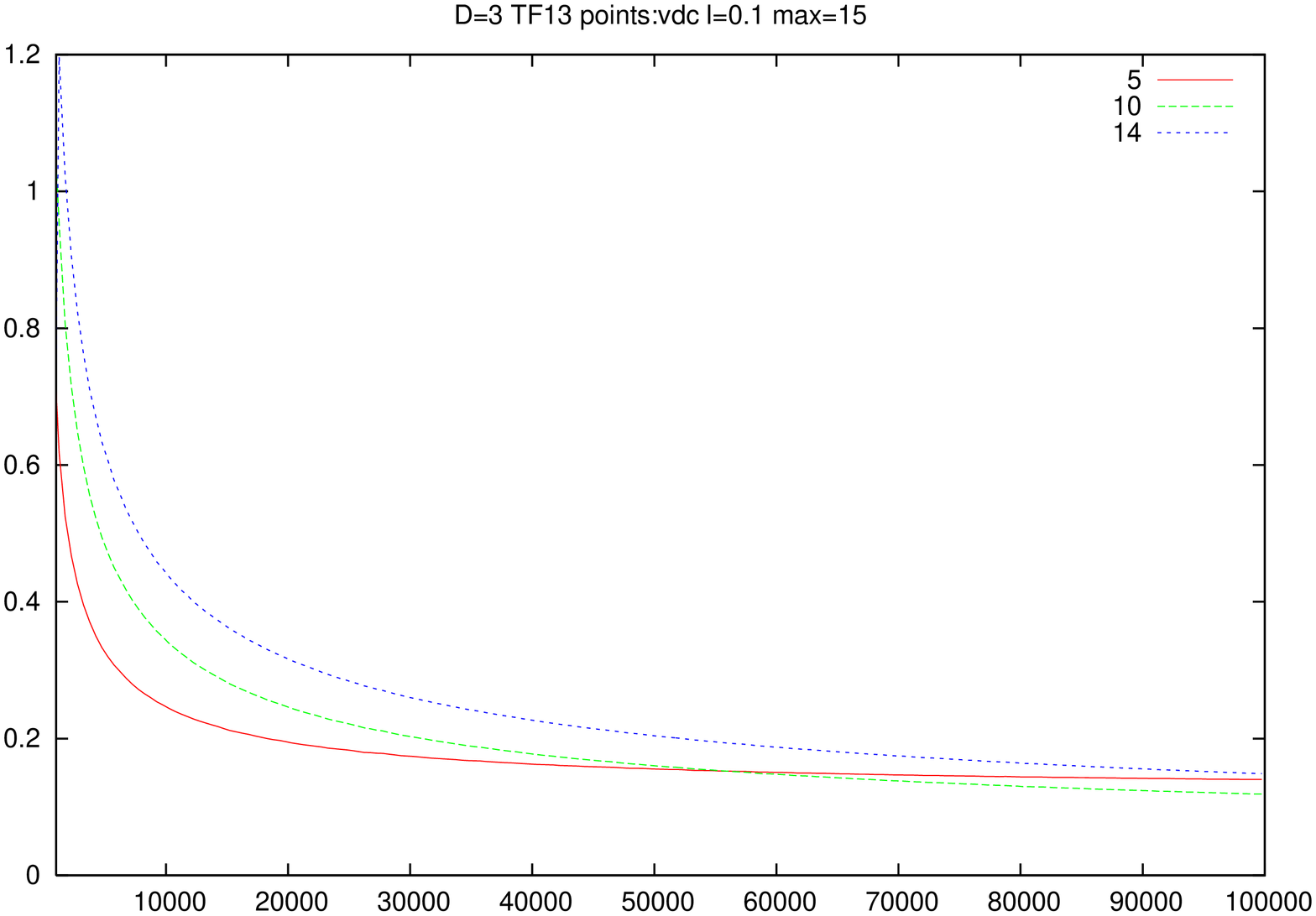, angle=270, width=7cm}&
\epsfig{file=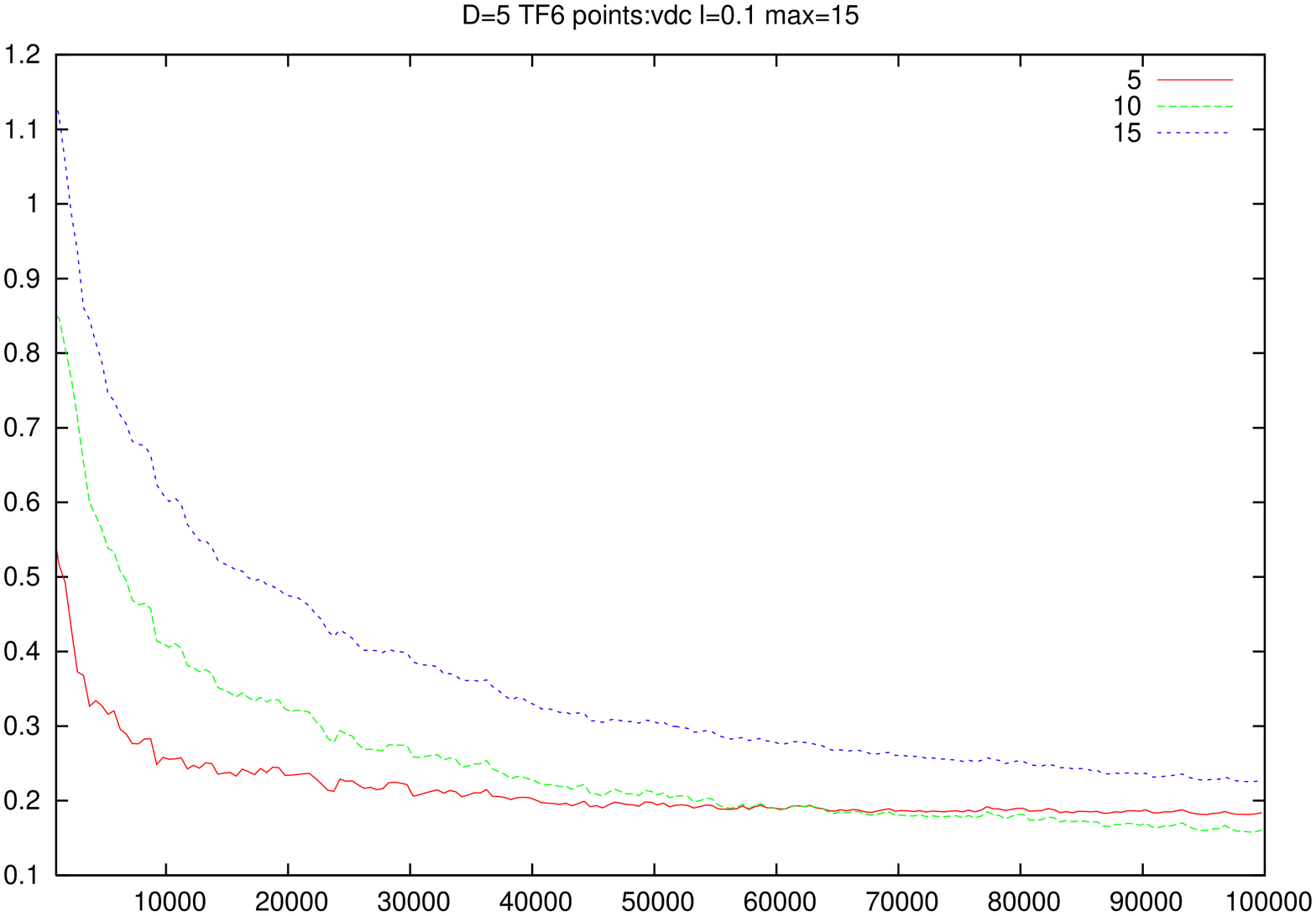, angle=270, width=7cm} 
\end{tabular}
\caption[ratio]
{
The ratio of different quasi estimators with the classical estimator for $d=3$ and TF13 (left)
and $d=5$ TF6 (right). In both cases {\tt Van der Corput} point-sets were used.
}
\label{ratio}
\end{minipage}

\end{figure}

The plots of fig.\ref{ratio} show the ratio of $E^{q5}_2$ , $E^{q10}_2$ and $E_2^{q15}$ 
with $E_2$ for two particular cases.

\subsection{The box approximation}
\label{box_approximation}
There is a choice for the diaphony that allows us to perform the integrals of Eq.(\ref{fullformulae}) without
resorting to the saddle point approximation. That choice is 
\bq
\sigma_{\vn}^2=\frac{1}{M}\prod_{\mu=1..d}\theta(n^{\mu}\leq m)
\eq 
for some arbitrary $m$. The normalization (Eq.\ref{sigma}) determines $M=(2m+1)^D-1$. 

This diaphony includes only a finite number of modes, all of which are equally weighted. It can be seen as an approximation
to the Jacobi diaphony since for small $\lambda$ the latter gives $\sigma_{\vn}\approx 1$ for $|\vn|\leq n_c$
and $\sigma_{\vn}\approx 0$ for $|\vn|\geq n_c$ where $n_c$ is determined implicitly 
by the value of the Jacobi diaphony. 
The diaphony can be evaluated as a quadratic function on the point-set from
\bq
S=\frac{1}{N}\sum_{\vn}\sigma_{\vn}^2|\sum_{i}\en(\vx)|^2=\frac{1}{NM}\sum_{|\vn|\leq m}|\sum_{i}\en(\vx)|^2
\equiv\frac{1}{NM}\sum_{i,j}\psi(\vx_i-\vx_j)
\eq
with
\bq
\psi(\vx_i-\vx_j)=-1+\prod_{\mu=1}^{D}\frac{sin\left( (2m+1)\pi (x_i^{\mu}-x_j^{\mu}) \right)}{sin\left( \pi (x_i^{\mu}-x_j^{\mu})\right)}
\eq	
The distribution of point-sets with a particular value for $s$ is then found by explicitly
performing the $z$-integrals of Eq.(\ref{fullformulae}):
\bq
H(s)=\frac{K^{K} s^{K-1}}{\Gamma(K)}e^{-Ks}
\eq
with $K\equiv M/2$. 
Hence the correlation function is
\bq
F(s;\vx_i-\vx_j)=\frac{(1-s)}{M}\psi(\vx_i-\vx_j)
\eq
and the estimator\footnote{The use of the improved estimator of eq.\ref{final2} in 
the box approximation is prohibited by the quadruple sums that it would contain.} of Eq.(\ref{estimatorfinal}) becomes
\bq
E_2^{(q)}=\frac{1}{N^2}\sum f_i^2-\frac{1}{N^3}(\sum f_i)^2-\frac{1}{N^3}\frac{(1-s)}{M}\sum_{i,j}\psi(\vx_i-\vx_j)f_i f_j
\eq	
This form has the advantage of including all modes up to an arbitrary $m$ without much
effort, with the overhead, of course, of being quadratic in $N$. As $N$ grows beyond $10^5$ this becomes 
particularly impractical. For investigating purposes, however, this approach is useful in testing the behavior of
$E_2^{(q)}$ with more modes included (that is presumably the small $\lambda$ limit).

It is remarkable that in the limit $m\rightarrow \infty$ we have $\psi(\vx_i-\vx_j)=M\delta_{i,j}$, and this leads to $s=1$
\bq
E_2^{(q)}=\left( \frac{1}{N^2}\sum f_i^2-\frac{1}{N^3}(\sum f_i)^2 \right)
\eq
In that limit a good point-set would have to integrates well any mode using a finite number of points $N$. Since that is
impossible, all
point-sets will be evaluated as equally bad by the particular diaphony.

It is evident that one has to find an optimal value for $m$. In the following plot the estimator $E_2^{(q)}$ 
is shown for TF5 in 2 dimensions with different values for $m$ ranging
from $3$ to $30$.

\begin{figure}[htbp]
\epsfig{file=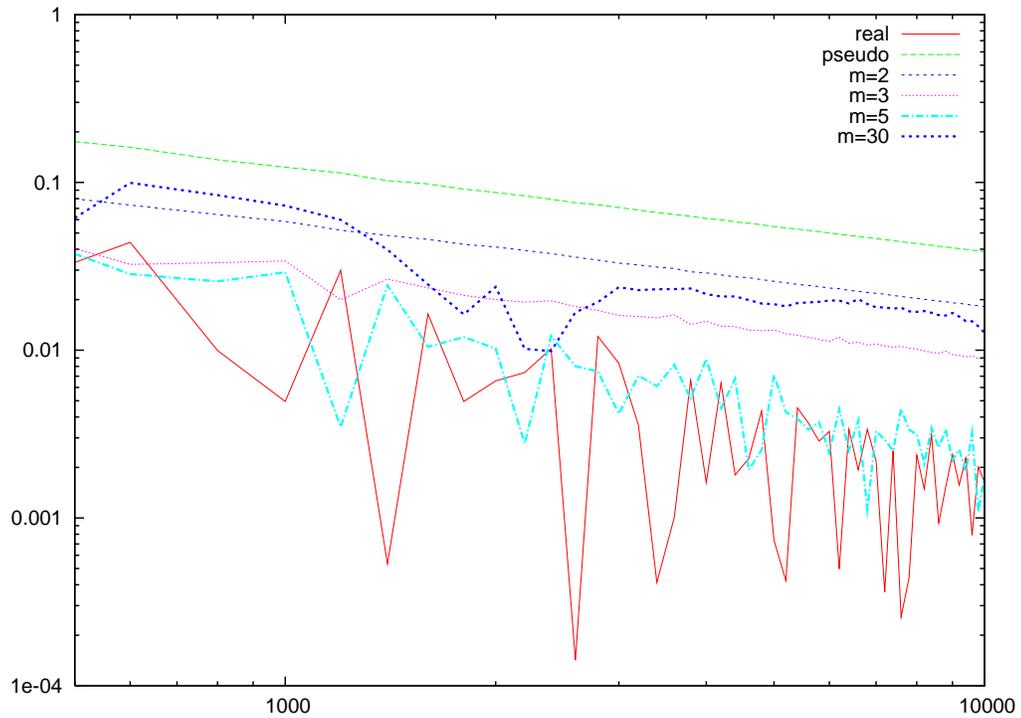, angle=270, width=14cm}
\caption[TF6 box, d=2]{
TF6, d=2 log-plot of the real error,
and then from top to bottom the classical estimate,
$E^{q}_2$ with $m=2$,
with $m=3$ and $m=5$.
The more modes one adds to the estimator the better it behaves. 
We also include the case $m=30$ (orange line),
to demonstrate that there is a turning point in $m$ 
above which the estimate becomes worse.
Note that $m=5$ means square length up to $2m^2=50$, 
much higher than $15$ that was our ceiling 
in the plots of the previous sections.
}
\end{figure}

\clearpage


\section{Concluding remarks}
\begin{itemize}

\item The use of \qmc\ point-sets in numerical integration achieves
a smaller error than the use of pseudo-random \mc\ point-sets. This advantage 
cannot be put in use without a reliable method for 
estimating the integration error.

\item The `classical', stochastic, error estimator relies 
on the assumption that
the points in the point-set are uncorrelated. When used with a \qmc\ point-set, this assumption
no longer holds. We saw that this leads to overestimating the error, thereby canceling any advantage gained
by using the \qmc\ point-set. 

\item An estimator of stochastic nature is still possible but the underlying
ensemble  can not be the ensemble of all point-sets. We advocate the use of
 the ensemble of point-sets with the same degree of uniformity, as measured
by a chosen diaphony. This approach leads to a prescription for a correlation function
and an estimator, without the use of any 
information on the particular point-set or integrand. 

\item The price to pay is the raise in the computational complexity
of the estimator from linear to quadratic in the 
number of points, which reflects the inclusion in the estimator of 
correlations between pairs of points.
Using properties of diaphonies one can revert to a complexity 
that is linear times the number of modes involved. 

\item The error estimator suggested in this paper is shown to 
perform better than the 'classical` error estimator, resulting in 
an estimate up to an order of magnitude smaller than the `classical' 
one.

\item The flexibility of the construction (reflected in the freedom to 
choose the precise diaphony and the number of modes included) allows one to trade accuracy
for computational cost. In computationally expensive
applications, the monitoring approach of section \ref{monitoring} could
be used to obtain an estimate that lies somewhere between the `classical'
and the quasi regime.
\end{itemize}

A number of further investigations have to be undertaken before implementing
\qmc\ in the demanding field of phase space integration in particle physics. We 
defer these and further testing of the error estimator suggested above, in realistic cases,
to further work.

\subsection*{acknowledgement}
We would like to thank dr.C.Papadopoulos for persistently reminding us
that a constant function can always be integrated with zero error.




\section*{Appendix A: Estimators by diagrammatics}

\subsection*{Diagrammatics for \qmc\ and \mc}
Our strategy for obtaining the form of the estimators is best described by
an example. Consider the triple sum
\bq
S_{p_1}S_{p_2}S_{p_3} \equiv \suml_{i,j,k=1}^N f_i^{p_1}f_j^{p_j}f_k^{p_3}\;\;.
\eq
In our approach we need to compute the expectation value of this object
including the first sub-leading order in $1/N$. It is given by
\bqa
\avg{S_{p_1}S_{p_2}S_{p_3}} &=& \fall{N}{3}\int f_i^{p_1}f_j^{p_j}f_k^{p_3}
\left( 1 - {1\over N}\left(
F_2(i,j) + F_2(i,k) + F_2(j,k)\right)\right)\nl && +
\fall{N}{2}\int \left(f_i^{p_1+p_2}f_k^{p_3} +
f_i^{p_1+p_3}f_j^{p_2} + f_i^{p_1}f_j^{p_2+p_3}\right) +
{\cal O}(N)\nl
&\approx& N^3\int f_i^{p_1}f_j^{p_j}f_k^{p_3}
- N^2\int f_i^{p_1}f_j^{p_j}f_k^{p_3}\left(\al_{ij}+\al_{ik}+\al_{jk}\right)
\nl &&+ N^2\int \left(f_i^{p_1+p_2}f_k^{p_3} +
f_i^{p_1+p_3}f_j^{p_2} + f_i^{p_1}f_j^{p_2+p_3}\right)\;\;,
\eqa
with implied integration over the subscripts. The sub-leading terms in the
expectation value are, therefore, obtained by either connecting any two
of the summands in the multiple sum $\Omega$ with a factor $-\al$, or
by contracting them. Now, any estimator $E$ consists of a linear combination
of terms like the above. Its variance, $\langle E^2\rangle - \avg{E}^2$,
contains both leading and sub-leading terms. The leading terms, however,
cancel completely, and so do the sub-leading terms coming from a 
connection/contraction {\em inside\/} one of the factors $E$. We arrive at the
following diagrammatic prescription. A sum of powers of $f$ will be represented
by a labeled dot, and a connection (including the $-\al$) by a link
between dots. For example,
\bq
\plaat{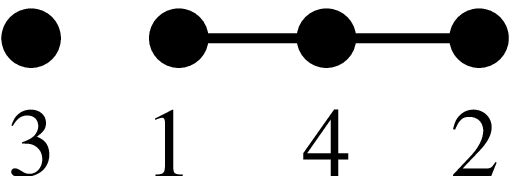}{2}{-0.5} = 
\suml_{i,j,k,l=1}^N f_i^3f_jf_k^4f_l^2\al_{jk}\al{kl}\;\;.
\eq
Now, suppose that the estimator $E$ is given as a linear combination
of {\em connected\/} diagrams. The estimator of its variance is the given
by the {\em connected sub-leading\/} diagrams that can be obtained from
$E\times E$. The factors $1/N$ can be added in a straightforward manner:
each sum with $p$ different summing indices carries a factor $N^{-p}$,
and there is an additional overall factor $N^{1-2^k}$ in $E_{2^k}$.

\subsection*{Estimators for \qmc}
We apply the above considerations to the first estimators $E^{(q)}_{1,2,4}$ for
\qmc. Squaring and constructing the connected sub-leading diagrams, we
find
\bqa
E^{(q)}_1 &=& \plaat{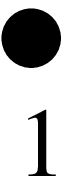}{0.3}{-0.3}\nl
E^{(q)}_2 &=& \plaat{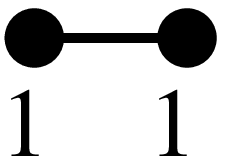}{1}{-0.3} + \plaat{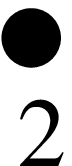}{0.3}{-0.3}\nl
E^{(q)}_4 &=& 4\;\plaat{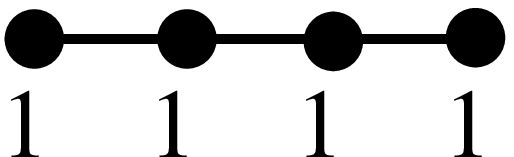}{2}{-0.3} + 4\;\plaat{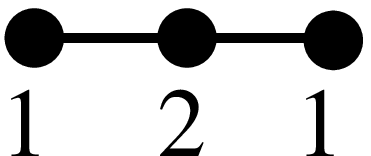}{1.5}{-0.3}
+ 4\;\plaat{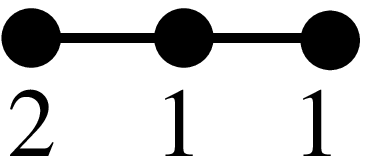}{1.5}{-0.3} + 4\;\plaat{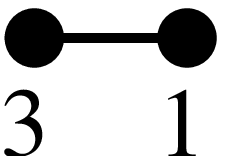}{1}{-0.3}\nl &&
+\;\plaat{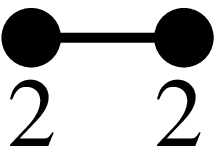}{1}{-0.3}\;+\;\plaat{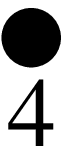}{0.4}{-0.3}\;\;.
\eqa
Upon insertion of the correct factors of $1/N$, we arrive precisely at
the estimators $E^{(q)}_{1,2,4}$ given in this paper.
The construction of $E^{(q)}_8$ is straightforward: at that order, tree diagrams
with branches develop. It may be worth noting that in this diagrammatic 
approach it becomes immediately clear that no diagrams with loops
(that is, occurrences of $\al_{jj}$, or $\al_{ij}\al_{ji}$, or
$\al_{ij}\al_{jk}\al_{ki}$, and so on) are possible to this order in $1/N$.

\subsection*{Estimators for \mc}
The MC estimators are of course precisely those of \qmc, with the replacement
$\al_{ij}\to1$. This means that the topology of the tree diagrams becomes
irrelevant, and we can feasibly go up to $E_{16}$. We find
\bq
E_{K} = {1\over N^{2K-1}}\suml_{s=0}^{K-1}E_{K,s}N^s\;\;\;,\;\;\;
K=1,2,4,8,16\;\;,
\eq
where the coefficients of the various powers of $N$ are given by
\bqa
E_{1,0} &=& S_1\;\;,\\
E_{2,0} &=& -S_1^2\;\;,\nl
E_{2,1} &=& S_2\;\;,\\
E_{4,0} &=& -4S_1^4\;\;,\nl
E_{4,1} &=& 8S_1^2S_2\;\;,\nl
E_{4,2} &=& -S_2^2-4S_1S_3\;\;,\nl
E_{4,3} &=& S_4\;\;,\\
E_{8,0} &=& -256S_1^8\;\;,\nl
E_{8,1} &=& 1024S_1^6S_2\;\;,\nl
E_{8,2} &=& -1152S_1^4S_2^2-512S_1^5S_3\;\;,\nl
E_{8,3} &=& 352S_1^2S_2^3+832S_1^3S_2S_3+224S_1^4S_4\;\;,\nl
E_{8,4} &=& -4S_2^4-224S_1S_2^2S_3-128S_1^2S_3^2
         -208S_1^2S_2S_4-96S_1^3S_5\;\;,\nl
E_{8,5} &=& 32S_2S_3^2+8S_2^2S_4+48S_1S_3S_4
         +48S_1S_2S_5+32S_1^2S_6\;\;,\nl
E_{8,6} &=& -S_4^2-8S_3S_5-4S_2S_6-8S_1S_7\;\;,\nl
E_{8,7} &=& S_8\;\;,\\
E_{16,0} &=& -4194304S_1^{16}\;\;,\nl
E_{16,1} &=& 33554432S_1^{14}S_2\;\;,\nl
E_{16,2} &=& -104857600S_1^{12}S_2^2-16777216S_1^{13}S_3\;\;,\nl
E_{16,3} &=& 162922496S_1^{10}S_2^3+93585408S_1^{11}S_2S_3+
          7733248S_1^{12}S_4\;\;,\nl
E_{16,4} &=& -132579328S_1^8S_2^4-189530112S_1^9S_2^2S_3
         -20185088S_1^{10}S_3^2\nl&&-37552128S_1^{10}S_2S_4 
        -3538944S_1^{11}S_5\;\;,\nl
E_{16,5} &=& 54444032S_1^6S_2^5+172064768S_1^7S_2^3S_3+
         69861376S_1^8S_2S_3^2+63553536S_1^8S_2^2S_4\nl&&
         +15532032S_1^9S_3S_4+14942208S_1^9S_2S_5
         +1507328S_1^{10}S_6\;\;,\nl
E_{16,6} &=& -9806848S_1^4S_2^6-69660672S_1^5S_2^4S_3-
   77729792S_1^6S_2^2S_3^2-45197312S_1^6S_2^3S_4\nl&&-
   43855872S_1^7S_2S_3S_4-5931008S_1^8S_3S_5
   -21135360S_1^7S_2^2S_5\nl&&-622592S_1^9S_7
   -5357568S_1^8S_2S_6-8060928S_1^7S_3^3-2802688S_1^8S_4^2\;\;,\nl
E_{16,7} &=& 551936S_1^2S_2^7+14180352S_1^5S_2S_3^3+
   10500096S_1^3S_2^5S_3+32006144S_1^4S_2^3S_3^2\nl&&+
   12816384S_1^4S_2^4S_4+6193152S_1^6S_2S_4^2+
   36679680S_1^5S_2^2S_3S_4\nl&&+7016448S_1^6S_3^2S_4+
   13725696S_1^6S_2S_3S_5+11722752S_1^5S_2^3S_5\nl&&+
   2007040S_1^7S_4S_5+6072320S_1^6S_2^2S_6+
   1994752S_1^7S_3S_6\nl&&+250880S_1^8S_8+1798144S_1^7S_2S_7\;\;,\nl
E_{16,8} &=& -256S_2^8-6438912S_1^3S_2^2S_3^3-
   3819520S_1^2S_2^4S_3^2-366592S_1S_2^6S_3\nl&&-
   1046016S_1^2S_2^5S_4-3568128S_1^4S_2^2S_4^2-
   8730624S_1^4S_2S_3^2S_4\nl&&-2233344S_1^3S_2^4S_5
   -1807360S_1^5S_3S_4^2-9879552S_1^3S_2^3S_3S_4
   \nl&&-8638464S_1^4S_2^2S_3S_5-2035712S_1^5S_3^2S_5
   -342016S_1^6S_5^2-2471936S_1^4S_2^3S_6\nl&&-
   3492864S_1^5S_2S_4S_5-1542144S_1^5S_2^2S_7
   -570368S_1^6S_2S_8-618496S_1^6S_3S_7\nl&&-
   607232S_1^6S_4S_6-851968S_1^4S_3^4-96256S_1^7S_9
   -3602432S_1^5S_2S_3S_6\;\;,\nl
E_{16,9} &=& 542208S_1S_2^4S_3S_4+
   2359296S_1^2S_2^2S_3^2S_4+1404160S_1^3S_2S_3S_4^2\nl&&
   +1608704S_1^3S_2^2S_3S_6+514432S_1^2S_2^3S_4^2+
   924672S_1^4S_3S_4S_5\nl&&+765952S_1^4S_2S_4S_6+
   552960S_1^2S_2S_3^4+1405952S_1^2S_2^3S_3S_5\nl&&+
   1814528S_1^3S_2S_3^2S_5+540672S_1S_2^3S_3^3+
   1394688S_1^3S_2^2S_4S_5\nl&&+262656S_1^2S_2^4S_6+
   808960S_1^4S_2S_3S_7+90624S_1S_2^5S_5+
   575488S_1^3S_3^3S_4\nl&&+451584S_1^4S_2S_5^2+
   191488S_1^5S_5S_6+475136S_1^4S_3^2S_6+
   164864S_1^5S_4S_7\nl&&+422912S_1^3S_2^3S_7+
   179200S_1^5S_3S_8+343296S_1^4S_2^2S_8+
   167936S_1^5S_2S_9\nl&&+1024S_2^6S_4+33792S_1^6S_{10}
   +133760S_1^4S_4^3+60416S_2^5S_3^2\;\;,\nl
E_{16,10} &=& -174336S_1S_2^2S_3S_4^2-
   191488S_1S_2S_3^3S_4-49920S_1^2S_2S_4^3
   -120832S_1S_2^3S_3S_6\nl&&-290304S_1^2S_2S_3^2S_6
   -183936S_1^2S_2^2S_4S_6-242688S_1S_2^2S_3^2S_5
   -99328S_1S_2^3S_4S_5\nl&&-87040S_1^3S_4^2S_5
   -231936S_1^2S_2^2S_3S_7-134400S_1^3S_2S_4S_7
   -153728S_1^3S_2S_3S_8\nl&&-174080S_1^3S_2S_5S_6
   -64512S_2^3S_3^2S_4-105216S_1^2S_3^2S_4^2
   -172288S_1^3S_3S_4S_6\nl&&-29184S_2^4S_3S_5
   -100352S_1^2S_3^3S_5-105472S_1^3S_3S_5^2
   -111360S_1^2S_2^2S_5^2\nl&&-90112S_1^3S_3^2S_7
   -22528S_1S_2^4S_7-40000S_1^4S_4S_8-47104S_1^4S_5S_7
   \nl&&-68608S_1^3S_2^2S_9-49088S_1^2S_2^3S_8
   -47616S_1^4S_3S_9-42240S_1^4S_2S_{10}\nl&&-
   10752S_1^5S_{11}-1152S_2^4S_4^2-23552S_2^2S_3^4
   -512S_2^5S_6-23296S_1^4S_6^2\nl&&-456960S_1^2S_2S_3S_4S_5
   -16384S_1S_3^5\;\;,\nl
E_{16,11} &=& 3328S_2^3S_5^2+20352S_2^2S_3S_4S_5+
   16576S_1S_2S_4^2S_5+25088S_1S_3^2S_4S_5\nl&&+
   27136S_1S_2S_3S_5^2+4096S_3^4S_4+
   33024S_1^2S_3S_5S_6+26240S_1^2S_2S_3S_9
   \nl&&+32768S_1S_2S_3^2S_7+17536S_1S_2^2S_4S_7
   +5184S_1S_3S_4^3+24064S_1^2S_3S_4S_7\nl&&+
   25856S_1^2S_2S_5S_7+19456S_1S_2^2S_3S_8
   +16448S_1^2S_2S_4S_8+20224S_1S_2^2S_5S_6
   \nl&&+11392S_2S_3^2S_4^2+13440S_1^2S_2S_6^2+
   11328S_1^2S_4^2S_6+10240S_2S_3^3S_5\nl&&+
   16000S_1^2S_4S_5^2+224S_2^4S_8+
   11520S_2^2S_3^2S_6+832S_2^3S_4S_6+
   13312S_1S_3^3S_6\nl&&+13568S_1^2S_3^2S_8+
   9472S_1^3S_6S_7+6912S_2^3S_3S_7+
   9984S_1^3S_5S_8\nl&&+8832S_1^3S_2S_{11}+
   10368S_1^3S_3S_{10}+4224S_1S_2^3S_9+
   8576S_1^3S_4S_9\nl&&+10176S_1^2S_2^2S_{10}+
   352S_2^2S_4^3+46336S_1S_2S_3S_4S_6+
   3008S_1^4S_{12}\;\;,\nl
E_{16,12} &=& -768S_3^2S_5^2-
   2944S_2S_3S_5S_6-1024S_3S_4^2S_5-
   2208S_1S_2S_5S_8-1664S_1S_2S_4S_9\nl&&-
   1216S_2S_4S_5^2-3072S_1S_2S_3S_{10}-
   2944S_2S_3S_4S_7-3584S_1S_3S_5S_7\nl&&-
   2944S_1S_2S_6S_7-2016S_1S_3S_4S_8-
   3008S_1S_4S_5S_6-768S_1^2S_7^2-
   1024S_3^3S_7\nl&&-1536S_3^2S_4S_6-1472S_1^2S_6S_8-
   224S_2S_4^2S_6-1920S_1S_3S_6^2-
   1024S_1S_4^2S_7\nl&&-1472S_2S_3^2S_8-
   1408S_2^2S_5S_7-208S_2^2S_4S_8-
   960S_1S_2^2S_{11}-1792S_1^2S_3S_{11}\nl&&-
   1440S_1^2S_2S_{12}-1312S_1^2S_4S_{10}-
   1792S_1S_3^2S_9-1088S_2^2S_3S_9\nl&&-
   1856S_1^2S_5S_9-768S_1S_5^3-128S_2^2S_6^2-
   704S_1^3S_{13}-4S_4^4-96S_2^3S_{10}\;\;,\nl
E_{16,13} &=& 128S_2S_7^2+128S_5^2S_6+32S_4S_6^2+
   160S_3S_5S_8+48S_2S_6S_8+8S_4^2S_8\nl&&+
   256S_3S_6S_7+192S_4S_5S_7+128S_3^2S_{10}+
   224S_1S_5S_{10}+160S_2S_5S_9\nl&&+128S_3S_4S_9+
   192S_1S_6S_9+160S_1S_7S_8+224S_1S_3S_{12}+
   48S_2S_4S_{10}\nl&&+32S_2^2S_{12}+192S_2S_3S_{11}+
   128S_1S_4S_{11}+160S_1S_2S_{13}+128S_1^2S_{14}\;\;,\nl
E_{16,14} &=& -S_8^2-16S_5S_{11}-16S_7S_9-8S_6S_{10}\nl&&-
   4S_4S_{12}-16S_3S_{13}-16S_1S_{15}-8S_2S_{14}\;\;,\nl
E_{16,15} &=& S_{16}\;\;.
\eqa
The number of individual terms in each $E_K$ is that of the partitions
$\Pi(K)$ of $K$: $\Pi(1)=1$, $\Pi(2)=2$, $\Pi(4)=5$, $\Pi(8)=22$, and
$\Pi(16)=231$. Likewise, the number of terms in each $E_{K,s}$ is the
partition of $K$ into $(K-s)$ parts.
We have not extended our results to the fifth-order error
estimator with $K=32$ and $\Pi(32)=8349$, since already $E_8$ and $E_{16}$
are purely academic and we have included them only as an illustration of
the method.


\section*{Appendix B: The $O(\frac{1}{N^2})$ contribution to $G_p$}
The second order contribution to $G_p$ can be found by summing up $O(\frac{1}{N^2})$ terms coming from
\begin{enumerate}
\item the pure rings (containing only 2-point vertices)
\item the three graphs contributing to $G_p^{(1,2,3)}$ (containing one 4-vertex, two 3-vertices or two external points)
\item products of a pure ring and one of the three graphs above or two of the graphs above. 
\item the new graphs (containing one 6-vertex,one 5-vertex and one 3-vertex, two 4-vertices, one 4-vertex and two 
3-vertices, four 3-vertices, one 3-vertex and three external points, two 3-vertices and two external points or one 4-vertex and two external points)
\end{enumerate}
After a lengthy but straightforward calculation (involving some cancellations) we get
\begin{eqnarray*}
G_p^{(3)}&=\frac{G_0p}{N^2} &( \;\frac{2p-1}{4}K_3-\frac{1}{4}(K_2)^2-\frac{1}{2}K_2(x_i,x_j)\; ) \\
		& +\frac{G_0}{N^2} &( \;\; \frac{3}{8}K_5+\frac{1}{3}K_4-\frac{1}{32}K_3^2-\frac{1}{8}K_2K_1(x_i,x_j)\\
		& & -\frac{1}{2}K_3(x_i,x_j)-\frac{1}{12}L_{1,1,1}-\frac{1}{2}L_{2,1,1}-\frac{1}{4}L_{2,2,1}                 \\
		& & -\frac{1}{48}K_3L_{1,1,1}-\frac{1}{4}L_{3,1,1}+\frac{1}{8} K_1(x_i,x_j)^2+\frac{1}{2}Q_1(x_i,x_j,x_k)\\
		& & +\frac{1}{4}Q_2(x_i,x_j)+\frac{1}{2}Q_3(x_i,x_j,x_k)+\frac{1}{4}Q_4(x_i,x_j)\\
		& & +\frac{1}{6}L_{1,1,1}(x_i,x_j,x_k)+\frac{1}{24}L_{1,1,1}K_1(x_i,x_j)+\frac{1}{288}L_{1,1,1}^2\\
		& & +\frac{1}{48}M_1+\frac{1}{8}M_2+\frac{1}{24}M_3+\frac{1}{16}M_4)
\end{eqnarray*}
where 
\bq
K_{a,b,\ldots}\equiv\sum_{1,2,\ldots} \rho_1^a\rho_2^b\ldots
\eq
\bq
K_a(x_i,x_j)\equiv\sum^{\prime}_{i,j}\sum_1\rho_1^a e_{\vn_1}(x_i)e^*_{\vn}(x_j) 
\eq
\bq
L_{a,b,c}\equiv \sum_{1,2,3} \rho_1^a  \rho_2^b \rho_3^c \delta_{1+2+3}
\eq
\bq
Q_1(x_i,x_j,x_k)\equiv\sum^{\prime}_{i,j,k}\sum_{1,2}\rho_1\rho_2 e_{\vn_1}(x_i)e^*_{\vn_1}(x_j)e_{\vn_2}(x_j)e^*_{\vn_2}(x_k)
\eq
\bq
Q_2(x_i,x_j)\equiv\sum^{\prime}_{i,j}\sum_{1,2}\rho_1\rho_2e_{\vn_1}(x_i)e_{\vn_1}^*(x_j)e_{\vn_2}^*(x_i)e_{\vn_2}(x_i)
\eq
\bq
Q_3(x_i,x_j,x_k)\equiv\sum^i_{i,j,k}\sum_{1,2,3}\rho_1\rho_2e_{\vn_1}(x_i)e_{\vn_1}^*(x_j)e_{\vn_2}(x_j)e_{\vn_2}^*(x_k)
e_{\vn_3}(x_j)e_{\vn_3}^*(x_k)
\eq
\bq
Q_4(x_i,x_j)\equiv\sum_{i,j}\sum_{1,2}\rho_1^2\rho_{1+2}e_{\vn_1}(x_i)e_{\vn_1}^*(x_j)
\eq
\bq
L_{a,b,c}(x_i,x_j,x_k)\equiv\sum^{\prime}_{i,j,k}\sum_{1,2,3} \rho_1^a\rho_2^b \rho_3^c e_{\vn_1}(\vx_i)e_{\vn_2}(\vx_j)e_{\vn_3}(\vx_k)\delta_{1+2+3}
\label{ThreePointFunction}
\eq
\bq
Q_{a,b}(x_i,x_j,x_k)\equiv\sum^{\prime}_{i,j,k}\sum_{1,2} \rho_1^a\rho_2^b e_{\vn_1}(\vx_i)e^*_{\vn_1}(\vx_j)e_{\vn_2}(\vx_j)e^*_{\vn_2}(\vx_k)
\eq
\bq
M_1\equiv\sum_{1,2,3,4}\rho_1\rho_2\rho_3\rho_4\delta_{1+2+3+4}
\eq
\bq
M_2\equiv\sum_{1,2,3,4,5}\rho_1\rho_2\rho_3\rho_4\rho_5
\delta_{1+2-5}\delta_{3+4-1-2}\delta_{5-3-4}
\eq
\bq
M_3\equiv\sum_{1,2,3,4,5,6}\rho_1\rho_2\rho_3\rho_4\rho_5\rho_6\delta_{1-2-5}\delta_{2-3-6}\delta_{3-1-4}\delta_{4+5+6}
\eq
\bq
M_4\equiv\sum_{1,2,3,4,5,6}\rho_1\rho_2\rho_3\rho_4\rho_5\rho_6\delta_{1-2-5}\delta_{2-3-6}\delta_{3+6-4}\delta_{4-5-1}
\eq

with 
\bq
\rho_i\equiv\frac{2z\sigma^2_{\vn_i}}{1-2z\sigma^2_{\vn_i}}
\eq
and $\sum^i_{i,j,k,\ldots}=\sum_{\vx_i\neq\vx_j\neq\vx_k\ldots}$, $\sum_{1,2,\ldots}\equiv\sum_{\vn_1,\vn_2,\ldots}$ and
$\delta_{1+8-2+\ldots}\equiv\delta(\vn_1+\vn_8-\vn_2+\ldots)$.

\end{document}